\documentclass[12pt]{article}

\usepackage{amsmath}
\usepackage{enumerate}
\usepackage{amssymb}
\usepackage{array}

\usepackage{hyperref}

\usepackage{graphicx}
\usepackage{spath3}
\usepackage{color}
\usepackage{mathtools}
\usepackage{amsfonts}

\usepackage[utf8]{inputenc}

\usepackage{tikz}
\usetikzlibrary{knots,hobby,arrows.meta}
\usetikzlibrary{shapes,snakes,spath3,intersections}

\usepackage{tikz-3dplot}
\usepackage{pgf}

\usepackage{enumitem}

\usepackage{pst-knot}

\date{}

\numberwithin{equation}{section}

\title{Solitons of the intermediate long wave equation \\with residue source}
\author{Yuto Ishii\footnote{e-mail: ishii.yuto@st.kitasato-u.ac.jp}, Atsushi Nakamula\footnote{e-mail: nakamula@sci.kitasato-u.ac.jp}\hspace{2mm}{}\footnote{Corresponding author} and Soma Yoshida\footnote{e-mail: yoshida.soma@st.kitasato-u.ac.jp
}  \\
\\
  Department of Physics, School of Science, Kitasato University\\
  Sagamihara 252-0373, Japan
}

\begin{document}

\maketitle

\begin{abstract}
We revisit the bilinear formulation of the intermediate long-wave (ILW) equation by interpreting its principal-value integrals in the framework of Sato hyperfunctions. 
When the tau-functions have no zeros inside the relevant contours, the conventional source-free ILW equation is recovered. 
When zeros enter the contours, their residue contributions must instead be retained, leading naturally to a residue-extended ILW system with self-consistent sources determined by the tau-functions themselves.
We introduce a residue index that counts the zeros enclosed in each analytic strip and use it to classify the resulting soliton sectors. 
The conventional elevation-soliton belongs to the zero-residue sector, whereas depression-solitons and higher elevation branches generally carry non-zero residue indices. 
Exact one- and multisoliton tau-functions are constructed, including mixed elevation--depression configurations with factorised two-body interactions. 
Their residue indices are additive and remain invariant during collisions, provided that no zero crosses a contour boundary. 
A Fourier-space analysis shows that the soliton spectra decay exponentially at large wavenumbers, with a decay rate determined by the distance of the nearest complex singularity from the real axis. 
This rate vanishes as an odd-multiple boundary between adjacent residue sectors is approached, providing a direct spectral signature of the transition between them.
\end{abstract}

\newpage

\section{Introduction}

Nonlinear dispersive equations of $(1+1)$-dimensions of the form
\begin{align}\label{dispersive wave equation} 
u_t+2uu_x-L[u_{x}]=0,
\end{align}
where $L$ is a linear operator acting on the spatial variable $x$, arise widely as asymptotic models for weakly nonlinear, long-wavelength wave propagation in a variety of physical settings \cite{SautPhysicaD40,Whitham1974}.
Throughout this paper, the subscripts describe partial derivatives with respect to the variables.
Such equations are conveniently classified by the dispersion relation $\hat{p}(\xi)$ associated with $L$ through its Fourier multiplier, $\widehat{L[g]}(\xi)=\hat{p}(\xi)\hat{g}(\xi)$, where $\xi$ is the wavenumber; in this convention, $\hat{p}(\xi)$ is an even function of $\xi$ for each of the equations considered here.
Two well-known members of this family are the Korteweg-de Vries (KdV) equation, with local, i.e., polynomial dispersion $\hat{p}(\xi)\sim \xi^2$, and the Benjamin-Ono (BO) equation \cite{Benjamin1967,OnoJPSJ39}, with genuinely nonlocal dispersion $\hat{p}(\xi)\sim |\xi|$, describing long internal waves in fluids of shallow and infinite depth, respectively.

The intermediate long wave (ILW) equation \cite{JosephJPA1977} interpolates between these two regimes.
It describes a $(1+1)$-dimensional dispersive wave amplitude at the boundary of two-layer fluids with finite depth, and it is a unidirectional wave equation with weak nonlinearity of the form \eqref{dispersive wave equation} with 
\begin{align}\label{dispersion term of ILW}
L[u_{x}]=-\frac{1}{2\delta}\;\mathrm{P}\partial_x^2\int_{-\infty}^\infty\coth\frac{\pi}{2\delta}(y-x)\,u(y)dy,
\end{align}
where $\mathrm{P}$ denotes principal value integral.
The parameter $\delta$ represents relative depth of the upper and lower layer, and the shallow- and deep-water limits correspond to $\delta\to0$ and $\delta\to\infty$, respectively.
The dispersion relation of the ILW equation is defined by 
\begin{align}
    \hat{p}(\xi)=2\pi \xi \coth\delta\xi,
\end{align}
the dispersion term is determined in a convolution form with the inverse Fourier transform of $\hat{p}(\xi)$, that is
\begin{align}
    L[u_{x}]=\mathrm{P}\int_{-\infty}^\infty \Check{p}(y-x) u_{y}(y)dy
    =\mathrm{P}\frac{\partial}{\partial x}\int_{-\infty}^\infty \Check{p}(y-x)u(y)dy,
\end{align}
where
\begin{align}
    \Check{p}(x)=\frac{1}{2\pi}\int_{-\infty}^\infty 2\pi \xi \coth\delta\xi \;e^{-ix\xi}\,d\xi
    =\frac{1}{2\delta}\frac{\partial}{\partial x}\coth \frac{\pi}{2\delta}x.
\end{align}
Hence, we find the dispersion term \eqref{dispersion term of ILW}.

Unlike a generic member of the family of equations \eqref{dispersive wave equation}, the ILW equation is completely integrable: it admits a Lax pair shown by Kodama, Ablowitz and Satsuma \cite{KodamaPRL46,KodamaJMP1982}, possesses an infinite hierarchy of conservation laws together with a B\"acklund transformation by Matsuno \cite{MatsunoJMP31:2904}.
Of particular relevance to the present work, the ILW equation can be recast into a Hirota bilinear form by Matsuno \cite{MatsunoPLA74} and Chen and Lee \cite{ChenPRL43}, from which exact $N$-soliton solutions can be constructed explicitly, including their determinant representations \cite{KodamaJMP1982}.
In terms of the tau-functions, the two-body interactions are encoded in closed-form phase-shift coefficients satisfying the factorisation property characteristic of an integrable system \cite{Hietarinta1988}.
This bilinear structure also provides a natural basis for the residue-extended ILW equation considered here.
When the conventional regularity condition on the tau-functions is relaxed, zeros of the tau-functions may enter the contours used to establish the equivalence between the bilinear equation and the source-free ILW equation.
The resulting contour integrals then acquire additional residue contributions. 
Retaining these contributions leads to an ILW equation with self-consistent residue sources, whose locations and evolution are determined by the zeros of the tau-functions themselves.

At the level of the bilinear system, the exact $N$-soliton determinant and its factorised scattering structure remain well defined.
The conventional elevation solitons belong to the zero-index sector, in which no residue source is present, whereas the higher elevation and depression branches belong to non-zero-index sectors carrying dynamically generated residue sources. 
Mixed elevation–depression solutions can consequently be interpreted as exact multisoliton solutions of this residue-extended bilinear ILW system.
This framework allows their profiles, regularity conditions, residue data, and pairwise scattering behaviour to be analysed in closed form, providing a controlled, non-perturbative setting in which to investigate how the nonlocal dispersion \eqref{dispersion term of ILW}, and particularly its finite-depth parameter $\delta$, governs the existence and character of solitary waves and the elevation–depression transition studied in Section \ref{section: elevation/depression solitons}.
In stratified fluids, depression-type internal solitary waves correspond to localized downward displacements of the density interface and have been observed in various oceanic environments, e.g., \cite{Helfrich2006,Orr2003}.
We expect that elucidating the physical significance of the self-consistent residue source will provide novel insights into nonlinear wave phenomena and their mathematical description.
Establishing complete integrability of the residue-extended equation itself would, however, require an independent construction of its Lax pair and conservation laws.

The ILW equation has also been extended in several directions. 
Xu \cite{Xu2012ILWsystems} derived a bidirectional ILW system describing the two-way propagation of long, weakly nonlinear internal waves along the interface between two fluid layers, using the asymptotic method developed by Bona, Lannes, and Saut \cite{BONA2008JMPA}. 
This system was shown to be well posed and to admit solitary-wave solutions, while the unidirectional ILW equation is recovered in an appropriate asymptotic regime. 
Numerical methods for computing solitary-wave solutions of such bidirectional systems, together with a detailed comparison of their performance, were subsequently developed by Bona, Durán, and Mitsotakis \cite{BonaDuranMitsotakis2021}.

In a different physical setting, Berntson, Langmann, and Lenells introduced the non-chiral ILW equation \cite{BerntsonLangmannLenells2020} to describe nonlinear edge-magnetoplasmon waves propagating in opposite directions along the two edges of a narrow quantum Hall system and interacting through inter-edge coupling. 
The resulting two-component system is completely integrable: it possesses a Lax pair, a Hirota bilinear form, and an infinite hierarchy of conservation laws, and its exact multisoliton solutions are governed by the hyperbolic Calogero--Moser--Sutherland model \cite{Berntson.Nonl35}.

These bidirectional and non-chiral generalisations enlarge the ILW dynamics by introducing an additional propagating field. 
The residue-extended ILW equation considered in the present work provides a complementary extension. 
It retains the scalar, unidirectional field of the original ILW equation but incorporates the residue contributions generated when zeros of the tau-functions enter the contours underlying the bilinear formulation.
This construction enlarges the admissible analytic sectors of the scalar bilinear system and allows higher elevation- and depression-solitons, as well as mixed elevation--depression multisoliton configurations, to be treated as exact solutions carrying self-consistent residue sources. 
It also suggests a possible connection between the non-zero-index sectors of the scalar bilinear system and the additional chiral degree of freedom in the non-chiral ILW equation.
The purpose of the present paper is therefore to investigate this residue extension systematically and to determine how the finite-depth parameter, the analytic structure of the tau-functions, and the associated residue data govern the existence and interactions of these additional solitary-wave branches.

This paper is organised as follows. 
Section \ref{section: bilinear form} reinterprets the transformation of the ILW equation into its bilinear form from the perspective of Sato hyperfunctions. 
We examine the additional contributions that arise when zeros of the tau-functions enter the contours used in the bilinearisation and show that retaining these contributions leads naturally to an ILW equation with self-consistent residue sources. 
This formulation distinguishes the conventional zero-index sector from the non-zero-index sectors and provides the basis for the higher elevation- and depression-solitons considered in the following section. 
Section \ref{section: elevation/depression solitons} then shows, for the one- and two-soliton solutions, that elevation- and depression-solitons, as well as their composite configurations, can indeed be constructed consistently within this residue-extended framework; for each composite two-soliton configuration, a Fourier-space analysis is carried out to confirm this consistency.
Finally, Section \ref{section:concluding remarks} summarises the paper and discusses future prospects.

\section{Bilinear form of the ILW equation}\label{section: bilinear form}

In this section, we reconsider the derivation of the bilinear form of the ILW equation, applying Sato's hyperfunction interpretation to the principal value integral and examining in detail the restrictions on the integrand, for the purpose of constructing depression-type solitons in addition to the conventionally known elevation-type soliton solutions of the ILW equation.

It is well-known that the ILW equation
\begin{align}\label{ILW}
u_t+2uu_x+\frac{1}{2\delta}\;\mathrm{P}\partial_x^2\int_{-\infty}^\infty\coth\frac{\pi}{2\delta}(y-x)\,u(y)dy=0,
\end{align}
can be recast as a bilinear differential equation \cite{MatsunoPLA74,ChenPRL43} 
\begin{align}\label{bilinear equation}
iD_t \;f^+\cdot f^-+D_x^2\;f^+\cdot f^-=0,
\end{align}
through a dependent variable transformation 
\begin{align}\label{dependent variable transf}
u(x,t)=-i\partial_x\log\frac{f^+}{f^-},
\end{align}
where $f^\pm$ are functions of independent real variables $x$ and $t$, and are complex conjugate each other.
The bilinear operators $D_t$ and $D_x$ are defined in the usual way \cite{HirotaPTP52}. 
The functions $f^\pm(x,t)$ are given through an analytic function $f(y,t)$ on the complex $y$-plane and real $t$ by
\begin{align}
 f^{\pm}(x,t):=f(x\pm i\delta,t).
\end{align}
In order to recast \eqref{ILW} into \eqref{bilinear equation}, we apply the following formula for the principal-value integral,
\begin{align}\label{PV integral}
&\mathrm{P}\int_{-\infty}^\infty\coth\frac{\pi}{2\delta}(y-x)\,\left(\partial_y\log\frac{f^{+}}{f^-}(y,t)\right)\;dy\nonumber\\
=&\mathrm{P}\int_{-\infty}^\infty\coth\frac{\pi}{2\delta}(y-x)\,\left(\frac{f_y^{+}}{f^+}(y,t)-\frac{f_y^{-}}{f^-}(y,t)\right)\;dy\nonumber\\
=&2i\delta\left(\frac{f_x^{+}}{f^+}(x,t)+\frac{f_x^{-}}{f^-}(x,t)\right)-2\pi iF(x,t),
\end{align}
where the integrand is assumed non-singular in the strip $-i\delta\leq\mathrm{Im} y\leq i\delta$ of the $y$-plane except for the first-order pole at $y=x\in\mathbb{R}$ from the $\coth$ funtion and the zeros of $f^\pm$.
The last term of the right-hand-side of \eqref{PV integral} is the contribution of residues from the zeros of $f^\pm$ inside the integration contours; if there is no poles it does not contribute to the principal-value integral and provides the conventional bilinear form of the ILW equation.
The reason of appearance of the first two terms in the right-hand-side of \eqref{PV integral} is essentially due to Cauchy's integral formula, and here we give a detailed derivation by using Sato's hyperfunction interpretation \cite{imai1992applied,morimoto1993introduction} of the principal-value integral.

\begin{figure}
    \centering

\tikzset{every picture/.style={line width=0.75pt}} 

\begin{tikzpicture}[x=0.75pt,y=0.75pt,yscale=-.9,xscale=.9]

\draw    (1.5,299.5) -- (650,300) ;
\draw [color={rgb, 255:red, 208; green, 2; blue, 27 }  ,draw opacity=1 ]   (17,65.5) -- (378,62.5) ;
\draw [color={rgb, 255:red, 208; green, 2; blue, 27 }  ,draw opacity=1 ]   (304,293.5) -- (534.99,291.76) ;
\draw  [draw opacity=0] (534.99,291.76) .. controls (535.14,280.43) and (542.45,271.29) .. (551.47,271.27) .. controls (560.6,271.24) and (568.03,280.56) .. (568.06,292.09) .. controls (568.07,292.55) and (568.06,293.01) .. (568.03,293.46) -- (551.53,292.13) -- cycle ; \draw  [color={rgb, 255:red, 208; green, 2; blue, 27 }  ,draw opacity=1 ] (534.99,291.76) .. controls (535.14,280.43) and (542.45,271.29) .. (551.47,271.27) .. controls (560.6,271.24) and (568.03,280.56) .. (568.06,292.09) .. controls (568.07,292.55) and (568.06,293.01) .. (568.03,293.46) ;  
\draw [color={rgb, 255:red, 208; green, 2; blue, 27 }  ,draw opacity=1 ]   (568.5,293.5) -- (635.5,293.5) ;
\draw  [draw opacity=0] (567.06,305.85) .. controls (566.96,315.88) and (559.72,324.01) .. (550.71,324.1) .. controls (541.65,324.19) and (534.2,316.11) .. (533.99,306) -- (550.52,305.6) -- cycle ; \draw  [color={rgb, 255:red, 208; green, 2; blue, 27 }  ,draw opacity=1 ] (567.06,305.85) .. controls (566.96,315.88) and (559.72,324.01) .. (550.71,324.1) .. controls (541.65,324.19) and (534.2,316.11) .. (533.99,306) ;  
\draw [color={rgb, 255:red, 208; green, 2; blue, 27 }  ,draw opacity=1 ]   (567.06,305.85) -- (634,305.5) ;
\draw [color={rgb, 255:red, 208; green, 2; blue, 27 }  ,draw opacity=1 ]   (17,292.5) -- (302,293.49) ;
\draw [shift={(304,293.5)}, rotate = 180.2] [color={rgb, 255:red, 208; green, 2; blue, 27 }  ,draw opacity=1 ][line width=0.75]    (10.93,-3.29) .. controls (6.95,-1.4) and (3.31,-0.3) .. (0,0) .. controls (3.31,0.3) and (6.95,1.4) .. (10.93,3.29)   ;
\draw [color={rgb, 255:red, 208; green, 2; blue, 27 }  ,draw opacity=1 ]   (636,61.5) -- (380,62.49) ;
\draw [shift={(378,62.5)}, rotate = 359.78] [color={rgb, 255:red, 208; green, 2; blue, 27 }  ,draw opacity=1 ][line width=0.75]    (10.93,-3.29) .. controls (6.95,-1.4) and (3.31,-0.3) .. (0,0) .. controls (3.31,0.3) and (6.95,1.4) .. (10.93,3.29)   ;
\draw [color={rgb, 255:red, 208; green, 2; blue, 27 }  ,draw opacity=1 ]   (16,304.5) -- (306,306.98) ;
\draw [shift={(308,307)}, rotate = 180.49] [color={rgb, 255:red, 208; green, 2; blue, 27 }  ,draw opacity=1 ][line width=0.75]    (10.93,-3.29) .. controls (6.95,-1.4) and (3.31,-0.3) .. (0,0) .. controls (3.31,0.3) and (6.95,1.4) .. (10.93,3.29)   ;
\draw   (615.97,42.98) -- (595.97,42.96) -- (596,22) ;
\draw [color={rgb, 255:red, 208; green, 2; blue, 27 }  ,draw opacity=1 ] [dash pattern={on 0.84pt off 2.51pt}]  (17,65.5) -- (15,309.5) ;
\draw    (350,27) -- (350,370) ;
\draw  [draw opacity=0] (424.3,232.45) .. controls (402.83,244.3) and (374.52,251.5) .. (343.5,251.5) .. controls (276.4,251.5) and (222,217.81) .. (222,176.25) .. controls (222,134.69) and (276.4,101) .. (343.5,101) .. controls (388.87,101) and (428.43,116.4) .. (449.3,139.22) -- (343.5,176.25) -- cycle ; \draw  [color={rgb, 255:red, 208; green, 2; blue, 27 }  ,draw opacity=1 ] (424.3,232.45) .. controls (402.83,244.3) and (374.52,251.5) .. (343.5,251.5) .. controls (276.4,251.5) and (222,217.81) .. (222,176.25) .. controls (222,134.69) and (276.4,101) .. (343.5,101) .. controls (388.87,101) and (428.43,116.4) .. (449.3,139.22) ;  
\draw [color={rgb, 255:red, 208; green, 2; blue, 27 }  ,draw opacity=1 ]   (424.3,232.45) -- (454.37,217.7) ;
\draw [shift={(456.17,216.82)}, rotate = 153.87] [color={rgb, 255:red, 208; green, 2; blue, 27 }  ,draw opacity=1 ][line width=0.75]    (10.93,-3.29) .. controls (6.95,-1.4) and (3.31,-0.3) .. (0,0) .. controls (3.31,0.3) and (6.95,1.4) .. (10.93,3.29)   ;
\draw [color={rgb, 255:red, 208; green, 2; blue, 27 }  ,draw opacity=1 ]   (308,307) -- (533.99,306) ;
\draw [color={rgb, 255:red, 208; green, 2; blue, 27 }  ,draw opacity=1 ] [dash pattern={on 0.84pt off 2.51pt}]  (636,61.5) -- (634,305.5) ;

\draw (676,292.4) node [anchor=north west][inner sep=0.75pt]    {$\mathbb{R}$};
\draw (648,57.4) node [anchor=north west][inner sep=0.75pt]    {$i\delta $};
\draw (540,292.4) node [anchor=north west][inner sep=0.75pt]    {$\times $};
\draw (602,20.4) node [anchor=north west][inner sep=0.75pt]    {$y$};
\draw (49,266.4) node [anchor=north west][inner sep=0.75pt]    {$C_{\mathrm{in}}^{+}$};
\draw (96,315.4) node [anchor=north west][inner sep=0.75pt]    {$C_{\mathrm{out}}^{+}$};
\draw (538,246.4) node [anchor=north west][inner sep=0.75pt]    {$C_{+}{}_{\varepsilon }$};
\draw (540,335.9) node [anchor=north west][inner sep=0.75pt]    {$C_{-\varepsilon }$};
\draw (447,73.4) node [anchor=north west][inner sep=0.75pt]    {$C_{\mathrm{in} ,\mathrm{out}}^{+}$};

\end{tikzpicture}

    \caption{Integration contour of $I^+_{\mathrm{in,out}}$ for the case of no zeros of $f^-$ inside}
    \label{fig:contour+}
\end{figure}
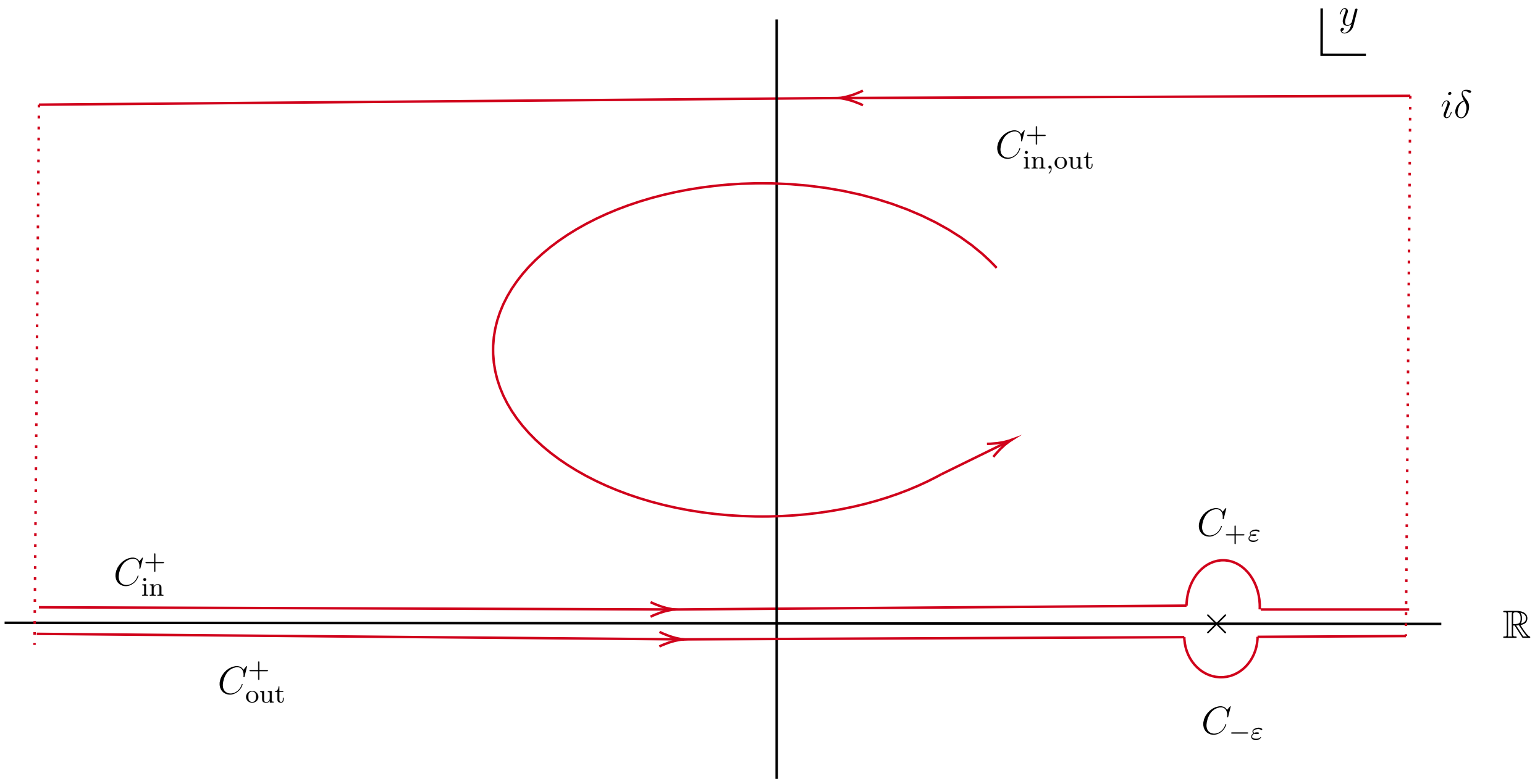

First, consider the following contour integrals on the complex $y$-plane,
\begin{align}\label{I+ in}
I^+_{\mathrm{in}}:=\oint_{C^+_{\mathrm{in}}}\coth\frac{\pi}{2\delta}(y-x)\;\frac{f_y^-}{f^-}(y)dy,\\
I^+_{\mathrm{out}}:=\oint_{C^+_{\mathrm{out}}}\coth\frac{\pi}{2\delta}(y-x)\;\frac{f_y^{-}}{f^-}(y)dy,\label{I+ out}
\end{align}
where $C^+_{\mathrm{in, out}}$ are anti-clockwise closed contours consisting of the real $y$-axis and the line $\mathrm{Im}\,y=\delta$ parallel to it, connected at $|\mathrm{Re}\; y|\to\infty$, respectively, as shown in Figure \ref{fig:contour+}.
The difference between these contours lies at the pole of the $\coth$ function $y=x$ on the real axis: at this point, $C^+_{\mathrm{in}}$ avoids the pole through the infinitesimal semicircle $C_{+\varepsilon}$, and $C^+_{\mathrm{out}}$ does through $C_{-\varepsilon}$, respecitively.
The contributions of the integral $I^+_{\mathrm{in,out}}$ from the vertical segments at $|\mathrm{Re}\;y|\to\infty$ are assumed to be constant.
This assumption is justified because the coth kernel tends to $x$-independent constants $\pm1$ as $|\mathrm{Re}\;y|\to\infty$, so that the integrals along the corresponding integration paths of finite length also approach constants, provided that $f^-_y/f^-$ remains bounded.
As we shall see in the next section, this condition is satisfied by the soliton solutions.

\vspace{10mm}

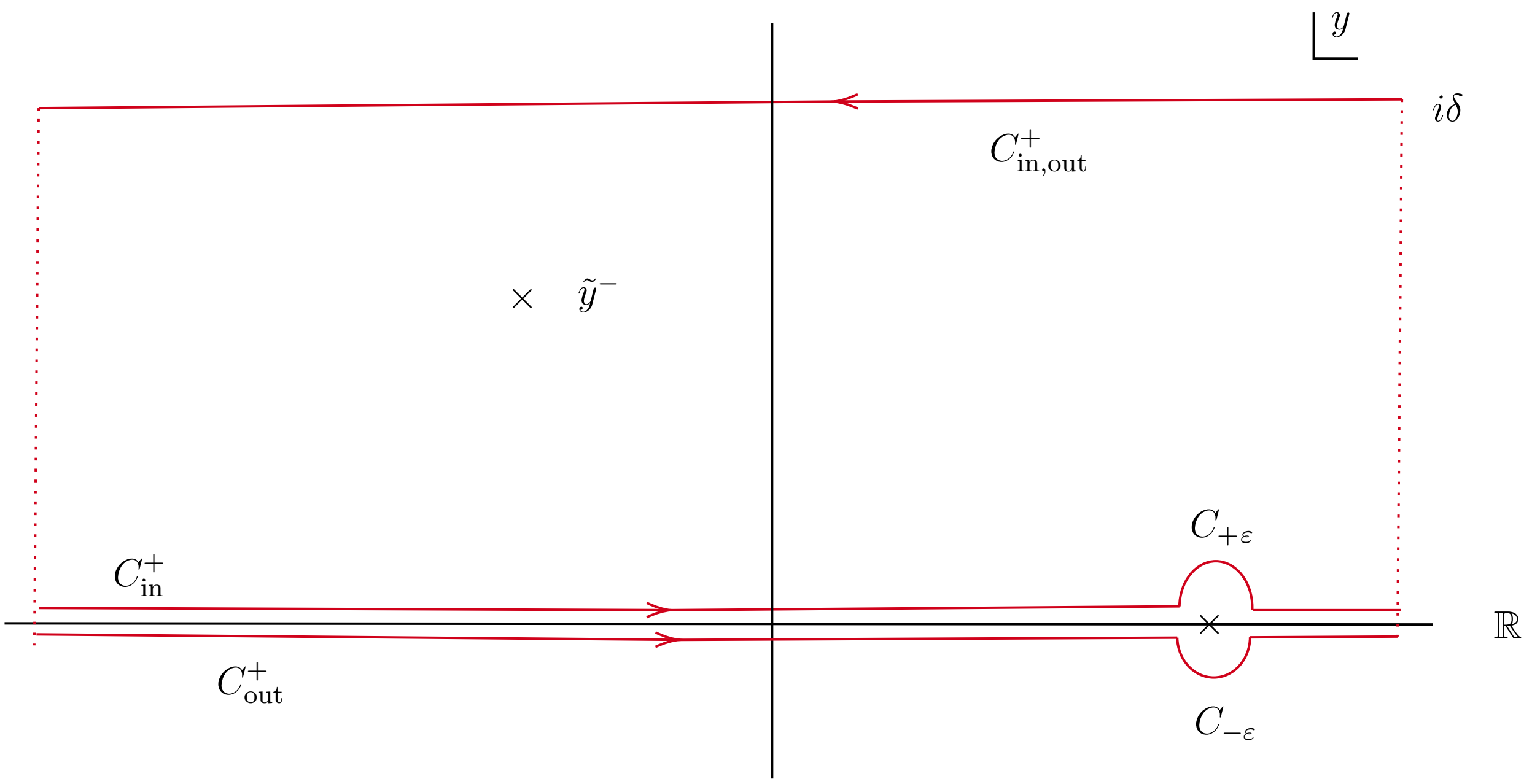
\begin{figure}
    \centering

\tikzset{every picture/.style={line width=0.75pt}} 

\begin{tikzpicture}[x=0.75pt,y=0.75pt,yscale=-.9,xscale=.9]

\draw    (1.5,299.5) -- (650,300) ;
\draw [color={rgb, 255:red, 208; green, 2; blue, 27 }  ,draw opacity=1 ]   (17,65.5) -- (378,62.5) ;
\draw [color={rgb, 255:red, 208; green, 2; blue, 27 }  ,draw opacity=1 ]   (304,293.5) -- (534.99,291.76) ;
\draw  [draw opacity=0] (534.99,291.76) .. controls (535.14,280.43) and (542.45,271.29) .. (551.47,271.27) .. controls (560.6,271.24) and (568.03,280.56) .. (568.06,292.09) .. controls (568.07,292.55) and (568.06,293.01) .. (568.03,293.46) -- (551.53,292.13) -- cycle ; \draw  [color={rgb, 255:red, 208; green, 2; blue, 27 }  ,draw opacity=1 ] (534.99,291.76) .. controls (535.14,280.43) and (542.45,271.29) .. (551.47,271.27) .. controls (560.6,271.24) and (568.03,280.56) .. (568.06,292.09) .. controls (568.07,292.55) and (568.06,293.01) .. (568.03,293.46) ;  
\draw [color={rgb, 255:red, 208; green, 2; blue, 27 }  ,draw opacity=1 ]   (568.5,293.5) -- (635.5,293.5) ;
\draw  [draw opacity=0] (567.06,305.85) .. controls (566.96,315.88) and (559.72,324.01) .. (550.71,324.1) .. controls (541.65,324.19) and (534.2,316.11) .. (533.99,306) -- (550.52,305.6) -- cycle ; \draw  [color={rgb, 255:red, 208; green, 2; blue, 27 }  ,draw opacity=1 ] (567.06,305.85) .. controls (566.96,315.88) and (559.72,324.01) .. (550.71,324.1) .. controls (541.65,324.19) and (534.2,316.11) .. (533.99,306) ;  
\draw [color={rgb, 255:red, 208; green, 2; blue, 27 }  ,draw opacity=1 ]   (567.06,305.85) -- (634,305.5) ;
\draw [color={rgb, 255:red, 208; green, 2; blue, 27 }  ,draw opacity=1 ]   (17,292.5) -- (302,293.49) ;
\draw [shift={(304,293.5)}, rotate = 180.2] [color={rgb, 255:red, 208; green, 2; blue, 27 }  ,draw opacity=1 ][line width=0.75]    (10.93,-3.29) .. controls (6.95,-1.4) and (3.31,-0.3) .. (0,0) .. controls (3.31,0.3) and (6.95,1.4) .. (10.93,3.29)   ;
\draw [color={rgb, 255:red, 208; green, 2; blue, 27 }  ,draw opacity=1 ]   (636,61.5) -- (380,62.49) ;
\draw [shift={(378,62.5)}, rotate = 359.78] [color={rgb, 255:red, 208; green, 2; blue, 27 }  ,draw opacity=1 ][line width=0.75]    (10.93,-3.29) .. controls (6.95,-1.4) and (3.31,-0.3) .. (0,0) .. controls (3.31,0.3) and (6.95,1.4) .. (10.93,3.29)   ;
\draw [color={rgb, 255:red, 208; green, 2; blue, 27 }  ,draw opacity=1 ]   (16,304.5) -- (306,306.98) ;
\draw [shift={(308,307)}, rotate = 180.49] [color={rgb, 255:red, 208; green, 2; blue, 27 }  ,draw opacity=1 ][line width=0.75]    (10.93,-3.29) .. controls (6.95,-1.4) and (3.31,-0.3) .. (0,0) .. controls (3.31,0.3) and (6.95,1.4) .. (10.93,3.29)   ;
\draw   (615.97,42.98) -- (595.97,42.96) -- (596,22) ;
\draw [color={rgb, 255:red, 208; green, 2; blue, 27 }  ,draw opacity=1 ] [dash pattern={on 0.84pt off 2.51pt}]  (17,65.5) -- (15,309.5) ;
\draw    (350,27) -- (350,370) ;
\draw [color={rgb, 255:red, 208; green, 2; blue, 27 }  ,draw opacity=1 ]   (308,307) -- (533.99,306) ;
\draw [color={rgb, 255:red, 208; green, 2; blue, 27 }  ,draw opacity=1 ] [dash pattern={on 0.84pt off 2.51pt}]  (636,61.5) -- (634,305.5) ;

\draw (676,292.4) node [anchor=north west][inner sep=0.75pt]    {$\mathbb{R}$};
\draw (648,57.4) node [anchor=north west][inner sep=0.75pt]    {$i\delta $};
\draw (540,292.4) node [anchor=north west][inner sep=0.75pt]    {$\times $};
\draw (602,20.4) node [anchor=north west][inner sep=0.75pt]    {$y$};
\draw (49,266.4) node [anchor=north west][inner sep=0.75pt]    {$C_{\mathrm{in}}^{+}$};
\draw (96,315.4) node [anchor=north west][inner sep=0.75pt]    {$C_{\mathrm{out}}^{+}$};
\draw (538,246.4) node [anchor=north west][inner sep=0.75pt]    {$C_{+}{}_{\varepsilon }$};
\draw (540,335.9) node [anchor=north west][inner sep=0.75pt]    {$C_{-\varepsilon }$};
\draw (447,73.4) node [anchor=north west][inner sep=0.75pt]    {$C_{\mathrm{in} ,\mathrm{out}}^{+}$};
\draw (228,144.4) node [anchor=north west][inner sep=0.75pt]    {$\times $};
\draw (260,139.4) node [anchor=north west][inner sep=0.75pt]    {$\tilde{y}^{-}$};

\end{tikzpicture}

    \caption{A pole of $f^-_y/f^-$ existing at $\tilde{y}^-$ inside of the contour of  $I^+_{\mathrm{in,out}}$}
\label{fig:contour with pole}
\end{figure}

In terms of Sato's hyperfunctions, a principal value integral passing through a singularity $x\in\mathbb{R}$ is defined as the average of the integrals along the real axis deformed by small semicircular detours above and below the singularity: we describe it as a schematic representation,
\begin{align}
\mathrm{P}\int_{-\infty}^\infty:=\frac{1}{2}\left(\int_{\mathbb{R}'}+\int_{C_{+\varepsilon}}\right)+\frac{1}{2}\left(\int_{\mathbb{R}'}+\int_{C_{-\varepsilon}}\right),
\end{align}
where $\mathbb{R}'=(-\infty,x-\varepsilon)\cup(x+\varepsilon,\infty)$, and $C_{\pm\varepsilon}$ are semicircular contours defined above.
It follows that,
\begin{align}
\mathrm{P}\int_{-\infty}^\infty\coth\frac{\pi}{2\delta}(y-x)\;\frac{f_y^-}{f^-}(y)dy=\int_{\mathbb{R}'}\coth\frac{\pi}{2\delta}(y-x)\;\frac{f_y^-}{f^-}(y)dy
+(\mbox{Integrals along $C_{\pm\varepsilon}$}),
\end{align}
where $\varepsilon\to0$ is understood.
The sum of the integrals along the semicircles $C_{\pm\varepsilon}$ vanishes as follows.
For $C_{+\varepsilon}$, substituting  $y=x+\varepsilon e^{i\theta}$, we find
\begin{align}\label{C +epsilon upper}
&\lim_{\varepsilon\to0}\int_{C_{+\varepsilon}}\coth\frac{\pi}{2\delta}(y-x)\,\frac{f_y^-}{f^-}(y)dy
=\lim_{\varepsilon\to0}\int_{C_{+\varepsilon}}\left(\frac{2\delta}{\pi}\frac{1}{y-x}+O(y-x)\right)\frac{f_y^{-}}{f^-}(y)dy\nonumber\\
=&\lim_{\varepsilon\to0}\int_\pi^0\left(\frac{2\delta}{\pi}\frac{1}{\varepsilon e^{i\theta}}+O(\varepsilon)\right)\frac{f_y^{-}}{f^-}(y)\Bigg|_{y=x+\varepsilon e^{i\theta}}i\varepsilon e^{i\theta}d\theta\nonumber\\
=&-2i\delta\frac{f_x^{-}}{f^-}(x),
\end{align}
and similarly for $C_{-\varepsilon}$,
\begin{align}\label{C -epsilon upper}
&\lim_{\varepsilon\to0}\int_{C_{-\varepsilon}}\coth\frac{\pi}{2\delta}(y-x)\,\frac{f_y^-}{f^-}(y)dy\nonumber\\
=&\lim_{\varepsilon\to0}\int_{-\pi}^0\left(\frac{2\delta}{\pi}\frac{1}{\varepsilon e^{i\theta}}+O(\varepsilon)\right)\frac{f_y^{-}}{f^-}(y)\Bigg|_{y=x+\varepsilon e^{i\theta}}i\varepsilon e^{i\theta}d\theta\nonumber\\
=&2i\delta\frac{f_x^{-}}{f^-}(x).
\end{align}
Therefore, the sum of \eqref{C +epsilon upper} and \eqref{C -epsilon upper} vanishes, and we observe that the summation of the contour integrals \eqref{I+ in} and \eqref{I+ out} becomes
\begin{align}\label{contour int on UHP}
\frac{1}{2}(I_{\mathrm{in}}^++I_{\mathrm{out}}^+)=\mathrm{P}\int_{-\infty}^\infty\coth\frac{\pi}{2\delta}(y-x)\;\frac{f_y^-}{f^-}(y)dy
-\int_{-\infty}^\infty\tanh\frac{\pi}{2\delta}(y-x)\,\frac{f_y}{f}(y)dy,
\end{align}
up to an additive constant arising from the integrals along the vertical segments as $\lvert\operatorname{Re} y\rvert\to\infty$.
Here the second term of the right-hand-side is the ordinary integral along the line of $\mathrm{Im} \;y=i\delta$, and the formula
\begin{align}\label{coth to tanh}
\coth\left(\frac{\pi}{2\delta}(y-x)\pm i\frac{\pi}{2}\right)=\tanh\frac{\pi}{2\delta}(y-x),
\end{align}
is used.
We now consider the case that $f^-$ has zeros inside of the contour $I^+_{\mathrm{in,out}}$; Figure \ref{fig:contour with pole} shows the case that $f^-$ has zero at $y=\tilde{y}^-$.
From the Cauchy integral formula, or the residue theorem, \eqref{I+ in} and \eqref{I+ out} are evaluated as
\begin{align}\label{evaluation I+ in}
&I^+_{\mathrm{in}}=2\pi i\operatorname{Res}[\tilde{y}^-](x),\\
&I^+_{\mathrm{out}}=4i\delta\frac{f_x^-}{f^-}(x)+2\pi i\operatorname{Res}[\tilde{y}^-](x),\label{evaluation I+ out}
\end{align}
where $\operatorname{Res}[\tilde{y}^-](x)$ denotes symbolically all of the contribution from the poles coming from the zeros of $f^-$, and it is a function of $x$ and $t$. 
Thus, \eqref{contour int on UHP} yields
\begin{align}\label{upper part of PV}
\mathrm{P}\int_{-\infty}^\infty\coth\frac{\pi}{2\delta}(y-x)\;\frac{f_y^-}{f^-}(y)dy
-\int_{-\infty}^\infty\tanh\frac{\pi}{2\delta}(y-x)\,\frac{f_y}{f}(y)dy
\nonumber\\
=2i\delta\frac{f_x^-}{f^-}(x)+2\pi i \operatorname{Res}[\tilde{y}^-](x),
\end{align}
up to an additive constant.

\begin{figure}
    \centering

\tikzset{every picture/.style={line width=0.75pt}} 

\begin{tikzpicture}[x=0.75pt,y=0.75pt,yscale=-.9,xscale=.9]

\draw [color={rgb, 255:red, 208; green, 2; blue, 27 }  ,draw opacity=1 ]   (48,324.5) -- (287,323.5) ;
\draw    (26,65) -- (648,65) ;
\draw [color={rgb, 255:red, 208; green, 2; blue, 27 }  ,draw opacity=1 ]   (306,73) -- (532.99,74) ;
\draw  [draw opacity=0] (531.99,57.77) .. controls (532.14,46.44) and (539.45,37.3) .. (548.47,37.28) .. controls (557.54,37.25) and (564.93,46.44) .. (565.06,57.86) -- (548.53,58.14) -- cycle ; \draw  [color={rgb, 255:red, 208; green, 2; blue, 27 }  ,draw opacity=1 ] (531.99,57.77) .. controls (532.14,46.44) and (539.45,37.3) .. (548.47,37.28) .. controls (557.54,37.25) and (564.93,46.44) .. (565.06,57.86) ;  
\draw [color={rgb, 255:red, 208; green, 2; blue, 27 }  ,draw opacity=1 ]   (567.06,73.85) -- (636,73) ;
\draw  [draw opacity=0] (566.06,73.25) .. controls (565.96,83.28) and (558.72,91.41) .. (549.71,91.5) .. controls (540.65,91.59) and (533.2,83.51) .. (532.99,73.4) -- (549.52,73) -- cycle ; \draw  [color={rgb, 255:red, 208; green, 2; blue, 27 }  ,draw opacity=1 ] (566.06,73.25) .. controls (565.96,83.28) and (558.72,91.41) .. (549.71,91.5) .. controls (540.65,91.59) and (533.2,83.51) .. (532.99,73.4) ;  
\draw [color={rgb, 255:red, 208; green, 2; blue, 27 }  ,draw opacity=1 ]   (48,57.5) -- (303,56.77) ;
\draw [shift={(305,56.77)}, rotate = 179.84] [color={rgb, 255:red, 208; green, 2; blue, 27 }  ,draw opacity=1 ][line width=0.75]    (10.93,-3.29) .. controls (6.95,-1.4) and (3.31,-0.3) .. (0,0) .. controls (3.31,0.3) and (6.95,1.4) .. (10.93,3.29)   ;
\draw [color={rgb, 255:red, 208; green, 2; blue, 27 }  ,draw opacity=1 ]   (48,72.5) -- (305,73) ;
\draw [shift={(307,73)}, rotate = 180.11] [color={rgb, 255:red, 208; green, 2; blue, 27 }  ,draw opacity=1 ][line width=0.75]    (10.93,-3.29) .. controls (6.95,-1.4) and (3.31,-0.3) .. (0,0) .. controls (3.31,0.3) and (6.95,1.4) .. (10.93,3.29)   ;
\draw [color={rgb, 255:red, 208; green, 2; blue, 27 }  ,draw opacity=1 ]   (635,321.5) -- (289,323.49) ;
\draw [shift={(287,323.5)}, rotate = 359.67] [color={rgb, 255:red, 208; green, 2; blue, 27 }  ,draw opacity=1 ][line width=0.75]    (10.93,-3.29) .. controls (6.95,-1.4) and (3.31,-0.3) .. (0,0) .. controls (3.31,0.3) and (6.95,1.4) .. (10.93,3.29)   ;
\draw   (615.97,42.98) -- (595.97,42.96) -- (596,22) ;
\draw [color={rgb, 255:red, 208; green, 2; blue, 27 }  ,draw opacity=1 ] [dash pattern={on 0.84pt off 2.51pt}]  (48,57.5) -- (48,324.5) ;
\draw [color={rgb, 255:red, 208; green, 2; blue, 27 }  ,draw opacity=1 ] [dash pattern={on 0.84pt off 2.51pt}]  (634,57.01) -- (636,188) ;
\draw [color={rgb, 255:red, 208; green, 2; blue, 27 }  ,draw opacity=1 ] [dash pattern={on 0.84pt off 2.51pt}]  (636,199.99) -- (636,314.99) ;
\draw    (350,27) -- (350,370) ;
\draw  [draw opacity=0] (502.46,220.48) .. controls (479.28,250.95) and (425.38,272.31) .. (362.59,272.31) .. controls (278.64,272.31) and (210.59,234.14) .. (210.59,187.06) .. controls (210.59,139.97) and (278.64,101.81) .. (362.59,101.81) .. controls (405.87,101.81) and (444.93,111.95) .. (472.61,128.24) -- (362.59,187.06) -- cycle ; \draw  [color={rgb, 255:red, 208; green, 2; blue, 27 }  ,draw opacity=1 ] (502.46,220.48) .. controls (479.28,250.95) and (425.38,272.31) .. (362.59,272.31) .. controls (278.64,272.31) and (210.59,234.14) .. (210.59,187.06) .. controls (210.59,139.97) and (278.64,101.81) .. (362.59,101.81) .. controls (405.87,101.81) and (444.93,111.95) .. (472.61,128.24) ;  
\draw [color={rgb, 255:red, 208; green, 2; blue, 27 }  ,draw opacity=1 ]   (472.61,128.24) -- (495.27,141.49) ;
\draw [shift={(497,142.5)}, rotate = 210.32] [color={rgb, 255:red, 208; green, 2; blue, 27 }  ,draw opacity=1 ][line width=0.75]    (10.93,-3.29) .. controls (6.95,-1.4) and (3.31,-0.3) .. (0,0) .. controls (3.31,0.3) and (6.95,1.4) .. (10.93,3.29)   ;
\draw [color={rgb, 255:red, 208; green, 2; blue, 27 }  ,draw opacity=1 ]   (305,56.77) -- (531.99,57.77) ;
\draw [color={rgb, 255:red, 208; green, 2; blue, 27 }  ,draw opacity=1 ]   (565.06,57.86) -- (634,57.01) ;

\draw (663,57.4) node [anchor=north west][inner sep=0.75pt]    {$\mathbb{R}$};
\draw (638,309.4) node [anchor=north west][inner sep=0.75pt]    {$-i\delta $};
\draw (539,57.4) node [anchor=north west][inner sep=0.75pt]    {$\times $};
\draw (602,20.4) node [anchor=north west][inner sep=0.75pt]    {$y$};
\draw (98,81.4) node [anchor=north west][inner sep=0.75pt]    {$C_{\mathrm{in}}^{-}$};
\draw (541,11.4) node [anchor=north west][inner sep=0.75pt]    {$C_{+\varepsilon }$};
\draw (539,96.9) node [anchor=north west][inner sep=0.75pt]    {$C_{-\varepsilon }$};
\draw (185,28.4) node [anchor=north west][inner sep=0.75pt]    {$C_{\mathrm{out}}^{-}$};
\draw (150,291.4) node [anchor=north west][inner sep=0.75pt]    {$C_{\mathrm{in} ,\mathrm{out}}^{-}$};

\end{tikzpicture}
    \caption{Integration contour of $I^-_{\mathrm{in,out}}$ for the case of no zeros of $f^+$ inside}
    \label{fig:contour-}
\end{figure}
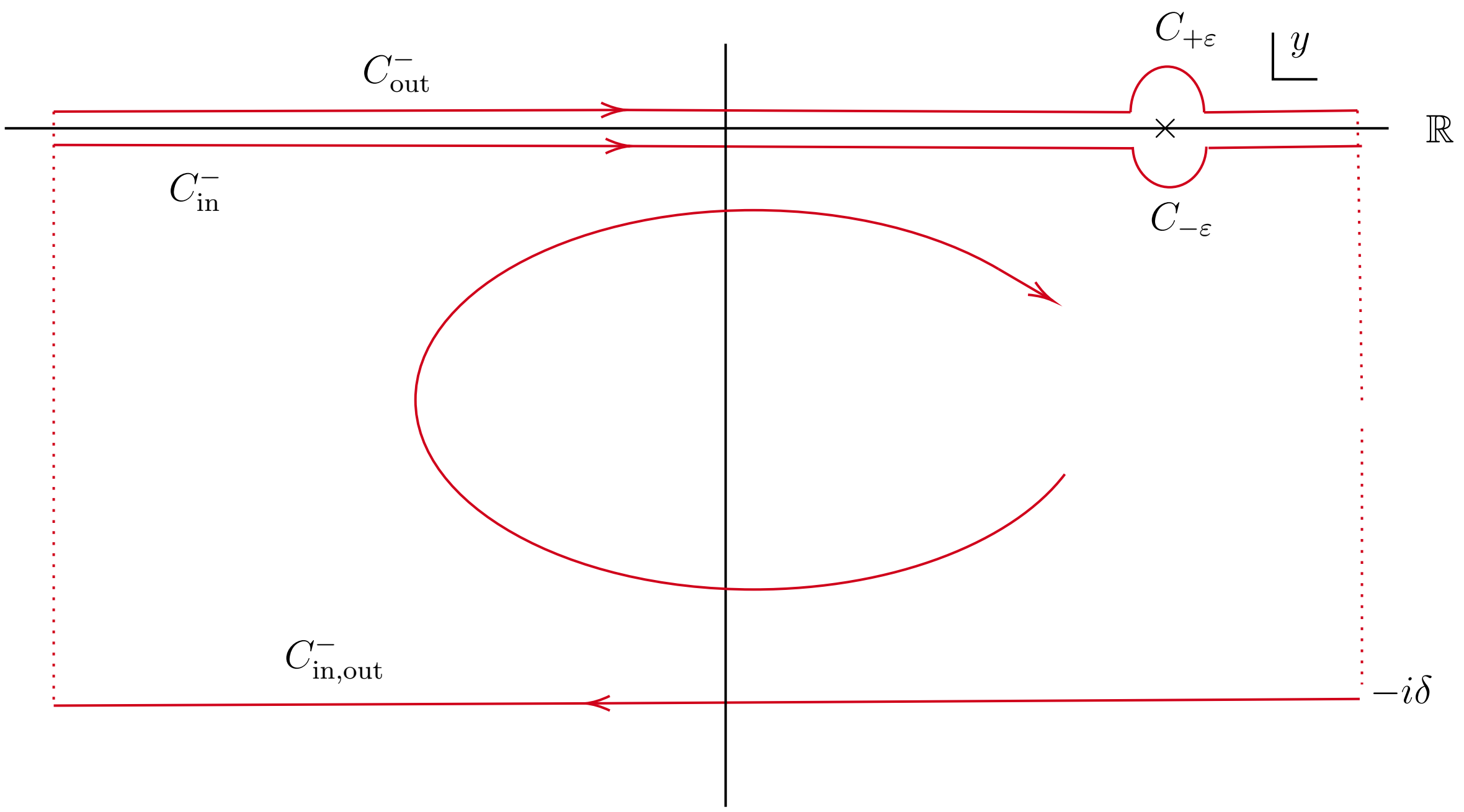

Next, consider the following pair of contour integrals,
\begin{align} \label{I- in}
I^-_{\mathrm{in}}:=\oint_{C^-_{\mathrm{in}}}\coth\frac{\pi}{2\delta}(y-x)\;\frac{f_y^+}{f^+}(y)dy,\\
I^-_{\mathrm{out}}:=\oint_{C^-_{\mathrm{out}}}\coth\frac{\pi}{2\delta}(y-x)\;\frac{f_y^{+}}{f^+}(y)dy.\label{I- out}
\end{align}
The contours $C^-_{\mathrm{in, out}}$ are defined analogously to $C^+_{\mathrm{in,out}}$ but are oriented clockwise in the lower half-plane.
They consist of segments along the real axis, the horizontal line $\mathrm{Im}\; y=-\delta$, and the vertical segments at $\lvert\operatorname{Re}y\rvert\to\infty$, as shown in Figure \ref{fig:contour-}.
Thus, we similarly observe that
\begin{align}\label{contour int on LHP}
\frac{1}{2}(I_{\mathrm{in}}^-+I_{\mathrm{out}}^-)=\mathrm{P}\int_{-\infty}^\infty\coth\frac{\pi}{2\delta}(y-x)\;\frac{f_y^+}{f^+}(y)dy
-\int_{-\infty}^\infty\tanh\frac{\pi}{2\delta}(y-x)\,\frac{f_y}{f}(y)dy,
\end{align}
and the contour integrals on the left-hand side are evaluated as
 \begin{align}\label{evaluation I- in}
&I^-_{\mathrm{in}}=-2\pi i\operatorname{Res}[\tilde{y}^+](x),\\
&I^-_{\mathrm{out}}=-4i\delta\frac{f_x^+}{f^+}(x)-2\pi i\operatorname{Res}[\tilde{y}^+](x),\label{evaluation I- out}
\end{align}
where the contribution of the zeros of $f^+$ at $y=\tilde{y}^+$'s are taking into account. 
Hence, \eqref{contour int on LHP} reads
\begin{align}\label{lower part of PV}
\mathrm{P}\int_{-\infty}^\infty\coth\frac{\pi}{2\delta}(y-x)\;\frac{f_y^+}{f^+}(y)dy
-\int_{-\infty}^\infty\tanh\frac{\pi}{2\delta}(y-x)\,\frac{f_y}{f}(y)dy\nonumber\\
=-2i\delta\frac{f_x^+}{f^+}(x)-2\pi i\operatorname{Res}[\tilde{y}^+](x),
\end{align}
up to an additive constant arising from the integrals along the vertical segments, as in the precedeng calculation.
Finally, eliminating the $\tanh$-integral term between \eqref{upper part of PV} and \eqref{lower part of PV}, we obtain the formula \eqref{PV integral}.

In deriving the principal-value integral formulas above, we have assumed that the logarithmic derivatives $f^-_y/f^-$ and $f^+_y/f^+$ have no poles on the contours $C^+_{\mathrm{in,out}}$ and $C^-_{\mathrm{in,out}}$, respectively. 
Poles lying strictly inside these contours are retained through their residue contributions.

We now derive \eqref{bilinear equation} by applying the dependent-variable transformation \eqref{dependent variable transf}.
The dispersion term in the ILW \eqref{ILW} can be written as
\begin{align}\label{dispersion term}
L[u_{x}]&=-\frac{1}{2\delta}\;\mathrm{P}\frac{\partial^2}{\partial x^2}\int_{-\infty}^\infty\coth\frac{\pi}{2\delta}(y-x)\,u(y)dy\nonumber\\
&=\frac{\partial}{\partial x}\left(\frac{\partial^2}{\partial x^2}\log f^+f^-\right)+\frac{\pi}{\delta}\frac{\partial^2}{\partial x^2}F(x,t),
\end{align}
where $F(x,t)$ denotes the sum of the residue contributios.

To characterise the analytic sector of a solution, we introduce the numbers of zeros of the tau-functions in the relevant strips. 
Let
\begin{align}
\Omega^-&=\{y\in\mathbb C\mid 0<\operatorname{Im}y<\delta\},\nonumber\\
\Omega^+&=\{y\in\mathbb C\mid -\delta<\operatorname{Im}y<0\},
\end{align}
and define
\begin{align}
\mathcal Z^-(t)
&=
\left\{
z\in\Omega^-\mid f^-(z,t)=0
\right\},
\nonumber\\
\mathcal Z^+(t)
&=
\left\{
z\in\Omega^+\mid f^+(z,t)=0
\right\},
\end{align}
namely, $\mathcal Z^\pm(t)$ denote sets of zeros of $f^\pm$ enclosed by the corresponding contours.
The corresponding zero counts, including multiplicities, are
\begin{align}\label{zero counts}
N^\pm(t)=
\sum_{z\in\mathcal{Z}^\pm(t)}
\operatorname{mult}_z(f^\pm).
\end{align}
Equivalently, provided that no zero lies on the boundary of the relevant strip, the argument principle gives
\begin{align}\label{zero counts argument principle}
N^\pm(t)
=
\frac{1}{2\pi i}
\oint_{\partial\Omega^\pm}
\frac{f_y^\pm}{f^\pm}(y,t)\,dy,
\end{align}
where the contours are counter-clockwise and the boundaries in \eqref{zero counts argument principle} are positively oriented.

For real solutions satisfying $f^+(z,t)=\overline{f^-(\overline z,t)}$,
the two zero sets occur in complex-conjugate pairs, and hence $N^+(t)=N^-(t)$.
We denote their common value by
\begin{align}\label{residue index}
\nu=N^+(t)=N^-(t)
\end{align}
and refer to it as the residue index of the solution.
The sector $\nu=0$ is the zero-residue sector corresponding to the conventional ILW equation, whereas $\nu>0$ characterises a non-zero-residue sector of the extended system. 
The residue index remains constant under time evolution unless a zero crosses one of the contour boundaries.
Then $F(x,t)$ is given by
\begin{align}\label{residue source}
F(x,t)
:=&
\sum_{\widetilde y^+\in\mathcal Z^+(t)}
\operatorname{Res}
\left[
\coth\frac{\pi}{2\delta}(y-x)
\frac{f_y^+}{f^+}(y,t),y=\widetilde y^+
\right]
\nonumber\\
+&
\sum_{\widetilde y^-\in\mathcal Z^-(t)}
\operatorname{Res}
\left[
\coth\frac{\pi}{2\delta}(y-x)
\frac{f_y^-}{f^-}(y,t),y=\widetilde y^-
\right]
\nonumber\\
=&
\sum_{\widetilde y^+\in\mathcal Z^+(t)}
\coth\frac{\pi}{2\delta}(\widetilde y^+-x)
+
\sum_{\widetilde y^-\in\mathcal Z^-(t)}
\coth\frac{\pi}{2\delta}(\widetilde y^--x).
\end{align}
Here we have asuumed that all the zeros of $f^\pm$ is simple, which can be confirmed for soliton solutions considered in the next section; the residue of the logarithmic derivative $f_y^\pm/f^\pm$ at each such zero is therefore equal to one.
For real solutions satisfying the appropriate conjugacy condition, the zeros in $\mathcal Z^+(t)$ and $\mathcal Z^-(t)$ occur in complex-conjugate pairs, and $F(x,t)$ is real-valued.
Any additive constant arising in the evaluation of the principal-value integral is eliminated by the differentiation with respect to $x$.

The nonlinear term reads
\begin{align}\label{nonlinear term}
2uu_x=(u^2)_x=-\frac{\partial}{\partial x}\left(\frac{\partial}{\partial x}\log\frac{f^+}{f^-}\right)^2,
\end{align}
and the sum of \eqref{dispersion term} and \eqref{nonlinear term} becomes
\begin{align}
2uu_x-L[u_{x}]=\frac{\partial}{\partial x}\left(-\frac{D_x^2 \;f^+\cdot f^-}{f^+f^-}\right)-\frac{\pi}{\delta}\frac{\partial^2}{\partial x^2}F(x,t).
\end{align}\label{nonlinear+dispersion}
Finally, the time-derivative term is
\begin{align}\label{t-derivative}
u_t=\frac{\partial}{\partial x}\left(\frac{-iD_t \;f^+\cdot f^-}{f^+f^-}\right).
\end{align}
If the sets $\mathcal{Z}^\pm$ are empty, then $F(x,t)=0$, and substitution of \eqref{dispersion term}, \eqref{nonlinear term}, and \eqref{t-derivative} into the ILW equation reduces in to the bilinear equation \eqref{bilinear equation}.
This is the conventional zero-residue sector.
However, for $F(x,t)\neq0$, the bilinear equation \eqref{bilinear equation} corresponds to ``Residue-extended ILW system",
\begin{align}\label{Residue-extended ILW}
    u_t+2uu_x+\frac{1}{2\delta}P\frac{\partial^2}{\partial x^2}\int_{-\infty}^\infty\coth\frac{\pi}{2\delta}(y-x)u(y)dy=-\frac{\pi}{\delta}\frac{\partial^2}{\partial x^2}F(x,t).
\end{align}
The term on the right-hand side is not an externally prescribed source. Rather, it is a self-consistent residue source determined by the zeros of the tau-functions through \eqref{residue source}.
We refer to \eqref{Residue-extended ILW}, together with the residue constraint \eqref{residue source}, as the residue-extended ILW system.
The conventional ILW equation is recovered as its zero-residue sector, $F(x,t)=0$.
When no ambiguity can arise, we refer to the residue-extended ILW system simply as the ILW system.

\section{Elevation and Depression Solitons of the ILW system}\label{section: elevation/depression solitons}

In this section, we construct explicit elevation- and depression-soliton solutions of the residue-extended ILW system and determine the constraints on their wavenumber parameters and associated residue sectors.
We also investigate two-soliton collisions in both configuration-space and Fourier-space and show that every combination of elevation- and depression-solitons is consistent with the contour formulation developed in the preceding section.
In particular, the zeros of tau-functions enclosed by the relevant contours generate precisely the self-consistent residue terms required by the residue-extended ILW equation.

\subsection{Soliton solutions}\label{subsection: soliton solutions}

Having obtained the bilinear equation \eqref{bilinear equation} and established its correspondence with the residue-extended ILW system, we can construct its soliton solutions directly.
The corresponding tau-function solutions were first derived in \cite{MatsunoPLA74,ChenPRL43}, and a determinant representation was subsequently given in \cite{KodamaJMP1982}. 
In these earlier studies, the wavenumber restriction
\begin{align}
    |k|<\frac{\pi}{\delta}
\end{align}
was imposed on each soliton as a consequence of the regularity condition on the tau-functions. 
This restriction ensures that no zeros of the tau-functions lie inside the relevant contours and therefore selects the conventional zero-residue sector corresponding to the source-free ILW equation.

Within the interpretation of the principal-value integral in terms of Sato hyperfunctions developed above, however, this restriction is not required for the solvability of the bilinear equation.
When it is relaxed, zeros of the tau-functions may enter the contours, but their contributions can be retained explicitly as the self-consistent residue source defined in \eqref{residue source}. 
The additional wavenumber ranges consequently describe non-zero-residue sectors of the extended system. 
As demonstrated below, these sectors contain higher elevation- and depression-soliton solutions, whose profiles and residue structures can be obtained explicitly.

\subsubsection{One-soliton solutions}
The bilinear equation \eqref{bilinear equation} can be solved by the standard Hirota method.
For the one-soliton solution, the tau-functions $f^\pm$ characterised by a wavenumber $k$, are given by
\begin{align}\label{1-soliton f: exp}
f^\pm(x,t)=1+e^{X^\pm(x,t)},
\end{align}
where $X^\pm(x,t)=k (x-c(k) \,t+\eta) \pm i\delta k$.
The bilinear equation determines the phase velocity to be $c(k)=-k\cot \delta k$, and $\eta\in\mathbb{R}$ is a constant.
In contrast to the KdV soliton and similar long-wave solitons, the phase velocity does not monotonically increasing with $|k|$, but piecewise monotonically increasing.

The contour prescription for the integrals $I^\pm_{\mathrm{in,out}}$, defined in \eqref{I+ in}, \eqref{I+ out}, \eqref{I- in}, and \eqref{I- out}, is well defined provided that the logarithmic derivatives $f_y^\pm/f^\pm$ have no poles on the corresponding contours.
Poles lying strictly inside the contours are allowed and contribute to the self-consistent residue source through \eqref{residue source}.
It also follows from \eqref{1-soliton f: exp} that the vertical segments of the integration contours at $|\mathrm{Re}\; y|\to\infty$ give constant contributions.
They are therefore eliminated by the $x$-derivatives acting on the principal value integrals.
This verifies the correspondence between the one-soliton tau-functions and the residue-extended ILW system.

For the one-soliton tau-functions, the only possible singularities of $f_y^\pm/f^\pm$ occur at the zeros of $f^\pm$.
From \eqref{1-soliton f: exp}, these zeros are located at
\begin{align}\label{zeros of one-soliton tau}
\widetilde{y}_n^{\,\mp}
&=
c(k)t-\eta
+i\left(
\pm\delta+\frac{(2n+1)\pi}{k}
\right),\qquad n\in\mathbb Z,
\end{align}
and all of which are simple.
A zero lies on the real-axis part of the corresponding contour precisely when
\begin{align}\label{wave number of odd multiple}
k=\frac{(2n+1)\pi}{\delta},
\qquad n\in\mathbb{Z}.
\end{align}
At these exceptional values, the singularity lies on the integration path and cannot be treated by the contour prescription used in deriving \eqref{PV integral}. 
These odd multiples must therefore be excluded.

The residue index \eqref{residue index} of the one-soliton solution can be evaluated explicitly from the zero locations \eqref{zeros of one-soliton tau}. 
For an admissible wavenumber $k$, it is given by floor function as
\begin{align}\label{one-soliton residue index}
\nu(k)
=
\left\lfloor
\frac{\delta|k|+\pi}{2\pi}
\right\rfloor,
\quad
\delta|k|\neq(2n+1)\pi.
\end{align}
Thus, the initial branch $0<\delta|k|<\pi$ has $\nu(k)=0$, so that no residue contribution is present and
$F(x,t)=0$.
This is the conventional zero-residue sector. 
For
$\nu(k)\geq1$, one or more complex-conjugate pairs of zeros lie strictly inside the upper and lower contours of $I^\pm_{\mathrm{in,out}}$ and generate the self-consistent residue source $F(x,t)$.
These ranges therefore belong to the non-zero-residue sectors of the extended system.

The non-zero even multiples $k=2n\pi/\delta$ must also be excluded, but for a different reason. 
Although they do not produce singularities on the integration paths, the phase velocity $c(k)=-k\cot\delta k$ is singular there, and the soliton solution degenerates.
The admissible non-degenerate wavenumbers therefore satisfy
\begin{align}\label{condition for 1-soliton wave number}
k\neq\frac{n\pi}{\delta},
\qquad n\in\mathbb Z.
\end{align}
The value $k=0$ is included among the excluded values for a separate reason: although the phase velocity has the finite limit $c(k)\longrightarrow-1/\delta$, the soliton amplitude vanishes $u\to0$.

We therefore find that the one-soliton solution of the ILW system is
\begin{align}\label{1-soliton}
u(x,t)&=-i\partial_x\log \frac{f^+}{f^-}\nonumber\\
&=-i\frac{k}{2}\left(\tanh\frac{X^+}{2}-\tanh\frac{X^-}{2}\right)\nonumber\\
&=\frac{k\sin \delta k}{\cosh k (x-c\, t+\eta)+\cos \delta k},
\end{align}
where $k\neq n\pi/\delta$.
Both the one-soliton profile \eqref{1-soliton} and its phase velocity are invariant under $k\to -k$, it is sufficient to consider the solutions in terms of $|k|$.

Except at the odd multiples \eqref{wave number of odd multiple}, the denominator in \eqref{1-soliton} is strictly positive. 
The sign of the solution is therefore determined by $k\sin(\delta k)$, giving
\begin{align}
\operatorname{sgn}u(x,t)
=
\begin{cases}
+1,
& 2n\pi<\delta|k|<(2n+1)\pi
\quad\mbox{(elevation branch)},\\[2mm]
-1,
& (2n+1)\pi<\delta|k|<(2n+2)\pi
\quad\mbox{(depression branch)},
\end{cases}
\quad
n\in\mathbb Z_{\geq0}.
\end{align}
Thus, elevation and depression branches occur alternately as $|k|$ increases.
Combining this classification with \eqref{one-soliton residue index}, we find that the initial elevation branch is the only zero-residue branch, while all the remaining elevation and depression branches have non-zero residue index.

For illustration, we show the residue contribution of a one-soliton in the first depression branch $\pi<\delta|k|<2\pi$.
In this case, a complex conjugate pair of zeros of $f^\pm$ appears in the strips $\Omega^\mp$, i.e., $\tilde{y}^-_{-1}\in\Omega^-$ and $\tilde{y}^+_0\in\Omega^+$ of \eqref{zeros of one-soliton tau}.
Thus, we find that \eqref{residue source} becomes
\begin{align}\label{residue F for 1st depression}
    F(x,t)=\coth\frac{\pi}{2\delta}(\tilde{y}^+_0-x)+\coth\frac{\pi}{2\delta}(\tilde{y}^-_{-1}-x)\nonumber\\
    =\frac{\displaystyle 2\sinh\frac{\pi}{\delta}(x-c(k)t+\eta)}{\displaystyle\cosh\frac{\pi}{\delta}(x-c(k)t+\eta)-\cos\pi(1-\frac{\pi}{\delta k})},
\end{align}
and the residue contribution is given by the second derivative of \eqref{residue F for 1st depression} with respect to $x$ as in the right-hand side of \eqref{Residue-extended ILW}.
Examples of one-soliton profiles in configuration space are shown in Figure 4 for $\delta=1$, with $k=2$ in the first elevation branch $0<\delta|k|<\pi$, and $k=4$ in the first depression branch $\pi<\delta|k|<2\pi$.

For fixed $k$, the depression branches disappear in the shallow-water, or KdV, limit: as $\delta\to0$, the first elevation branch expands to $0<|k|<\infty$.
The deep-water limit, by contrast, requires further consideration and will be discussed in the final section.

\begin{center}
\begin{figure}
    \centering
\includegraphics[width=70mm]{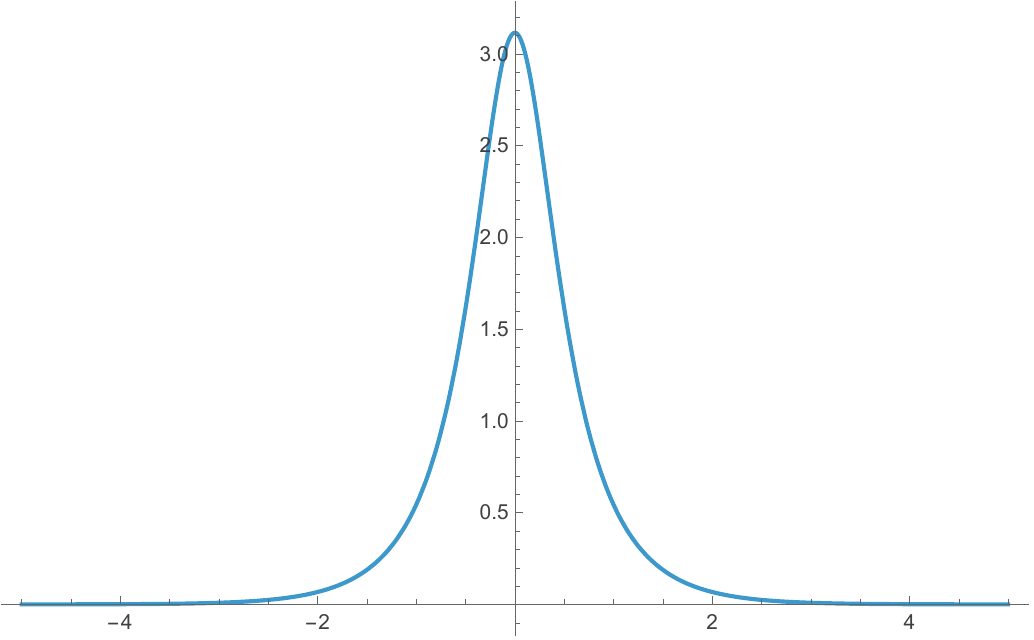}
\includegraphics[width=70mm]{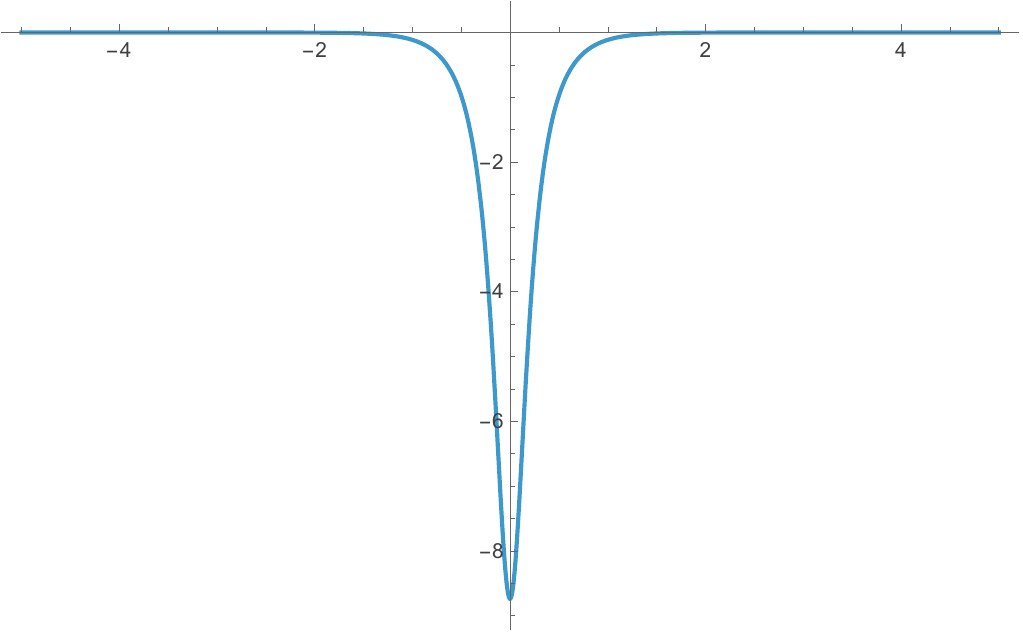}   
    \caption{Profile of one-solitons with $\delta=1$.
    Left: a soliton in the first elevation branch with $k=2$, Right: a soliton in the first depression branch with $k=4$.}
    \label{fig:1-solitons}
\end{figure}
 
\end{center}

\subsubsection{Two-soliton solutions}
\label{subsection: two-soliton}

We next consider the two-soliton solutions of the bilinear equation \eqref{bilinear equation}.
The standard Hirota construction gives
\begin{align}\label{2-soliton f}
f^\pm=1+e^{X_1^\pm}+e^{X_2^\pm}
+A_{12}e^{X_1^\pm+X_2^\pm},
\end{align}
where $
X_j^\pm=k_j\bigl(x-c_jt+\eta_j\bigr)\pm i\delta k_j$ and $c_j=-k_j\cot\delta k_j, \ j=1,2$.
The two-body interaction coefficient is
\begin{align}\label{2-soliton interaction}
A_{12}=
-\frac{
(c_1k_1-c_2k_2)\sin\delta(k_1-k_2)
+(k_1-k_2)^2\cos\delta(k_1-k_2)
}{
(c_1k_1+c_2k_2)\sin\delta(k_1+k_2)+(k_1+k_2)^2\cos\delta(k_1+k_2)}.
\end{align}

As in the one-soliton case, the contour prescription becomes singular if either wavenumber, say $k_1$, approaches an odd multiple of $\pi/\delta$:
\begin{align}
k_1\longrightarrow\frac{(2n+1)\pi}{\delta}.
\end{align}
At such a value, the phase velocity $c_1$ diverges.
For a fixed admissible value of $k_2$, the interaction coefficient has the limiting value
\begin{align}
A_{12}
&\longrightarrow
-\frac{
\sin\{(2n+1)\pi-\delta k_2\}
}{
\sin\{(2n+1)\pi+\delta k_2\}
}
=1.
\end{align}
Consequently, the tau-functions factorise as
\begin{align}
f^\pm
\longrightarrow
\bigl(1+e^{X_1^\pm}\bigr)
\bigl(1+e^{X_2^\pm}\bigr).
\end{align}
At the centre of the first soliton, $x-c_1t+\eta_1=0,$ we have $e^{X_1^\pm}=e^{\pm i(2n+1)\pi}=-1$.
Hence, $f^\pm=0$ at this point, independently of the admissible value of $k_2$.
The logarithmic derivatives $f_y^\pm/f^\pm$ therefore develop poles on the real-axis part of the corresponding contours.
These odd-multiple values of $k_1$, and similarly of $k_2$, must consequently be excluded from the two-soliton solutions.

Thus, the singularities associated with the individual soliton factors are avoided provided that $k_j\neq(2n+1)\pi/\delta,
\; j=1,2,\;n\in\mathbb Z$.
Zeros of tau-function lying strictly inside the contours are permitted and contribute to the self-consistent residue source.

The cases in which either $k_j$ is a non-zero even multiple of $\pi/\delta$ must be excluded for the same reason as in the one-soliton case: the corresponding phase velocity diverges. 
In this limit, the divergence of $c_j$ again gives $A_{12}\longrightarrow1$,
and the tau-functions factorise.
For example, when $k_1\longrightarrow 2n\pi/\delta,\; n\neq0,$
we have
\begin{align}
f^\pm
\longrightarrow
\bigl(1+e^{X_1^\pm}\bigr)
\bigl(1+e^{X_2^\pm}\bigr).
\end{align}
Moreover, $e^{X_1^+}=e^{X_1^-},$ because $e^{2i\delta k_1}=1$.
The first factor therefore cancels from the ratio $f^+/f^-$, and the physical field $u=-i\partial_x\log(f^+/f^-)$
reduces exactly to the one-soliton solution associated with $k_2$.
If $k_j=0$, the corresponding soliton amplitude vanishes and the solution likewise degenerates.
The wavenumber exclusions for a non-degenerate two-soliton solution are therefore
\begin{align}
k_j\neq\frac{n\pi}{\delta},
\qquad j=1,2,\qquad n\in\mathbb Z.
\end{align}
The contributions from the vertical parts of the contours as
$\lvert\operatorname{Re}y\rvert\to\infty$ are independent of $x$ and are consequently eliminated by the $x$-derivatives in the dependent-variable transformation.

Provided that the full two-soliton tau-functions have no zeros on the contour boundaries, their residue index is conserved during the collision.
In the asymptotic regions $t\to\pm\infty$, the tau-functions separate into one-soliton factors. 
The argument principle therefore gives
\begin{align}\label{two-soliton residue index}
\nu_2=
\nu(k_1)+\nu(k_2),
\end{align}
where $\nu(k)$ is defined by \eqref{one-soliton residue index}.
For solitons belonging to the initial elevation and depression branches, the elevation-elevation(EE), depression-elevation (DE), and depression-depression (DD) configurations consequently have residue indices $\nu_2=0,\, \nu_2=1$, and $\nu_2=2,$ respectively. 
Thus, the conventional EE solution belongs to the source-free sector, whereas the DE and DD solutions carry one and two complex-conjugate pairs of residue-generating zeros, respectively.
The individual zeros may move parallel to the real axis as time goes on, and exchange their association with the solitons during a collision, their total number remains unchanged.

\begin{figure}
    \begin{center}
        \includegraphics[width=70mm]{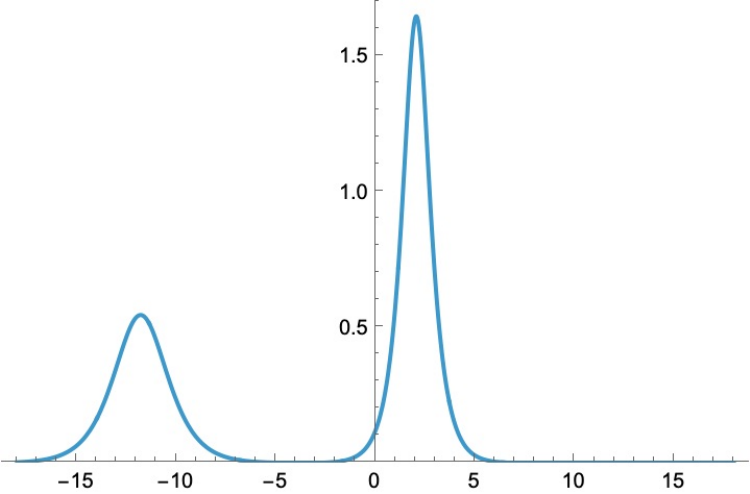}
        \includegraphics[width=70mm]{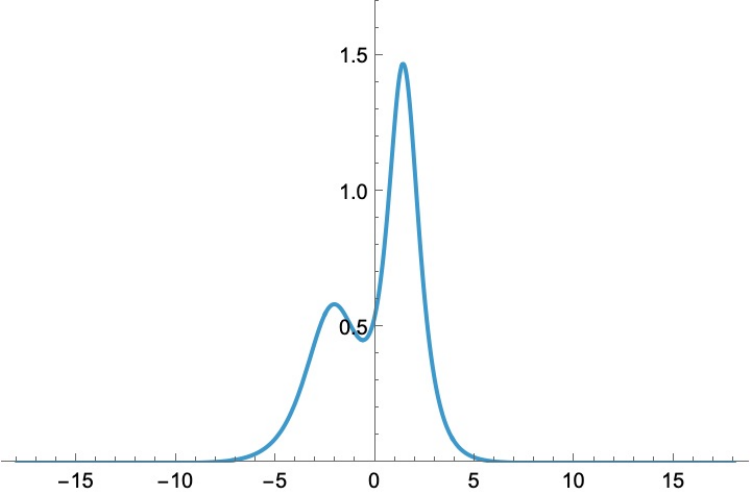}\\
        \includegraphics[width=70mm]{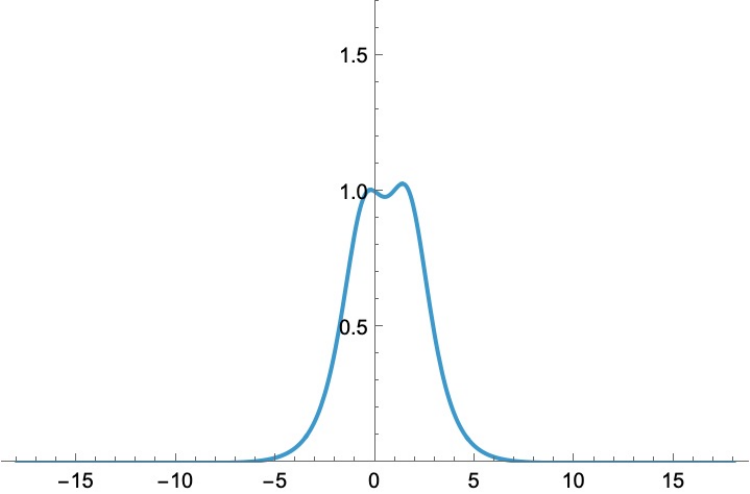}
        \includegraphics[width=70mm]{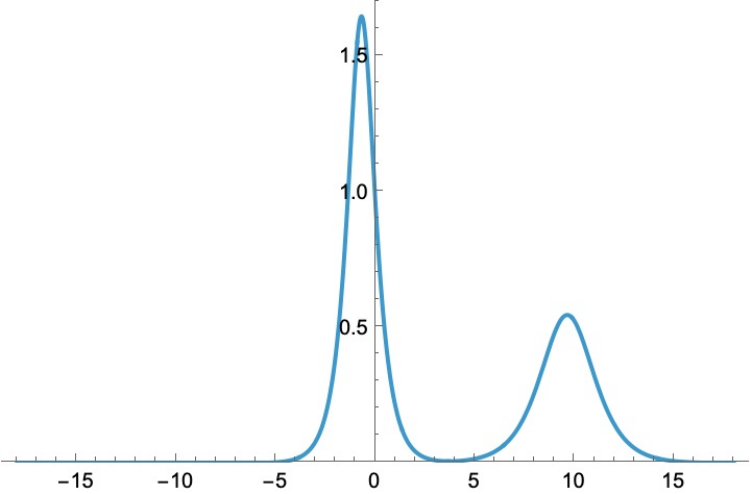}
               \caption{Two-solitons of type EE with $(k_1,k_2)=(1.0,-1.6)$ at $t=15.0$ (upper left), $t=0.0$ (upper right), $t=-2.5$ (lower left), and $t=-15.0$ (lower right).}
        \label{fig:EE1 config}
        \end{center}
    \end{figure}

\begin{figure}
    \begin{center}
        \includegraphics[width=70mm]{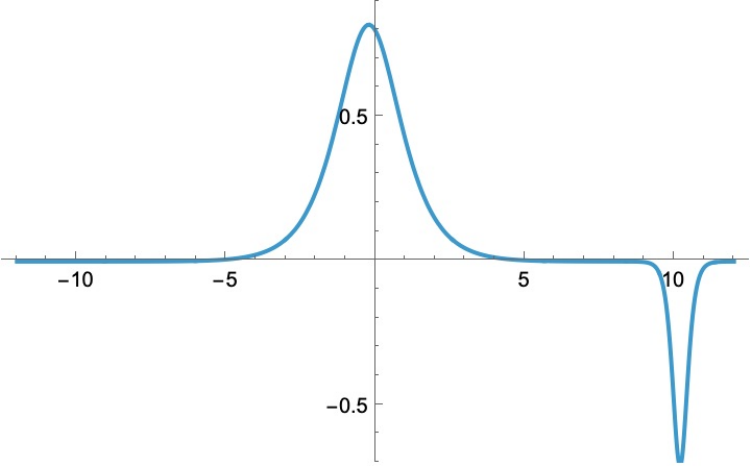}
        \includegraphics[width=70mm]{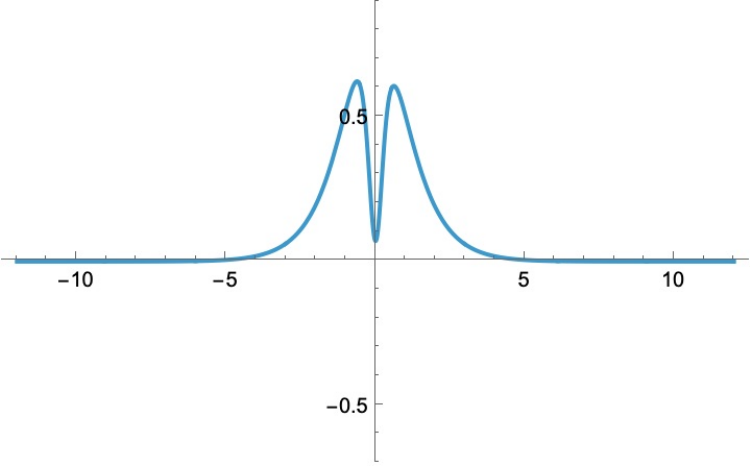}\\
        \includegraphics[width=70mm]{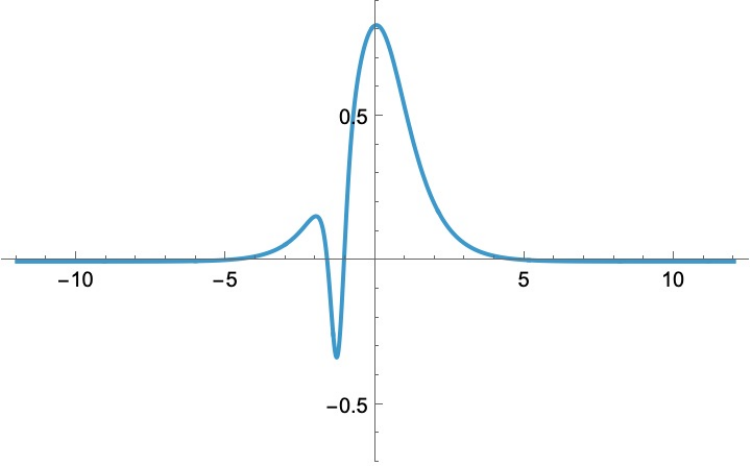}
        \includegraphics[width=70mm]{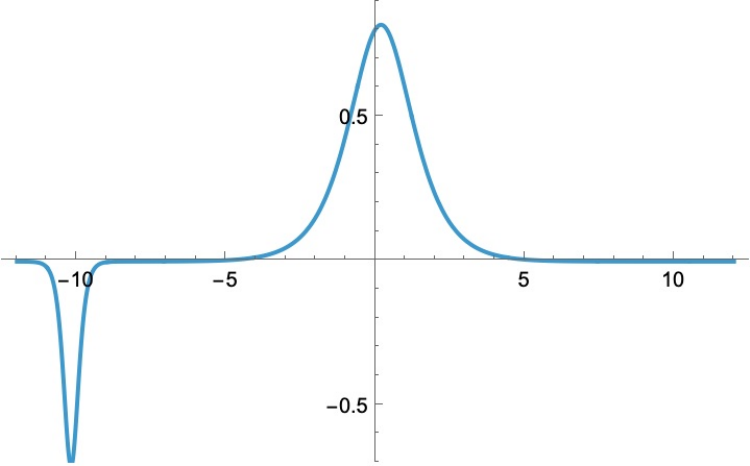}
               \caption{Two-solitons of type DE with $(k_1,k_2)=(1.2,-6.05)$ at $t=0.4$ (upper left), $t=0.0$ (upper right), $t=-0.05$ (lower left), and $t=-0.4$ (lower right).}
        \label{fig:DE1 config}
    \end{center}
\end{figure}

\begin{figure}
    \begin{center}
        \includegraphics[width=70mm]{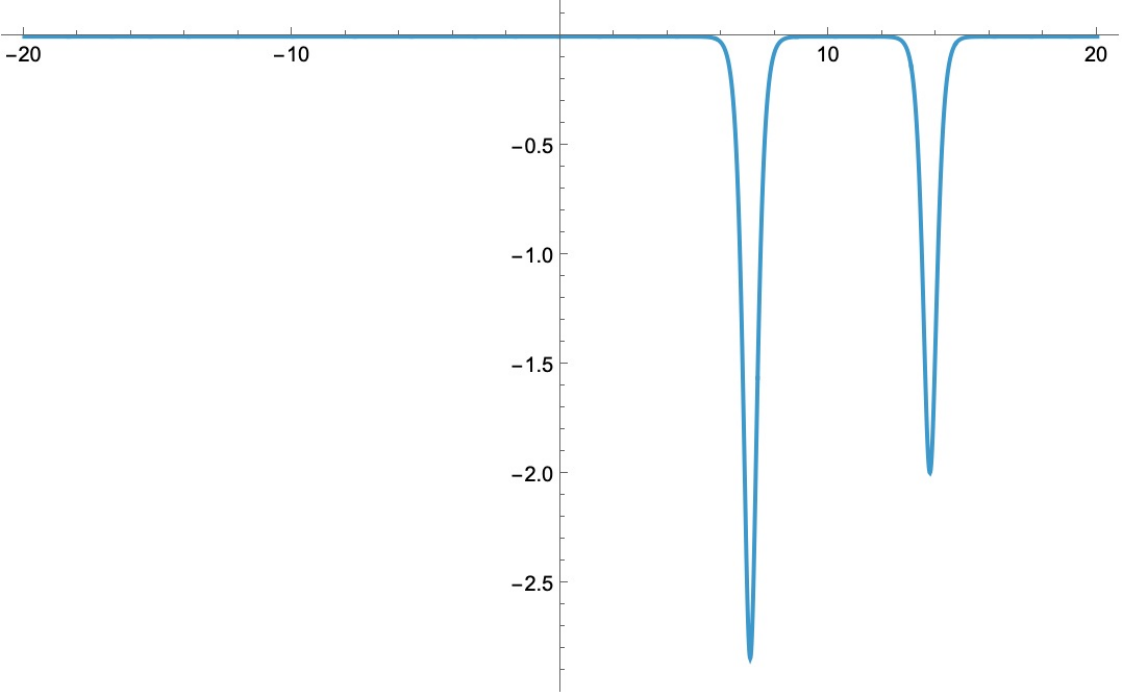}
        \includegraphics[width=70mm]{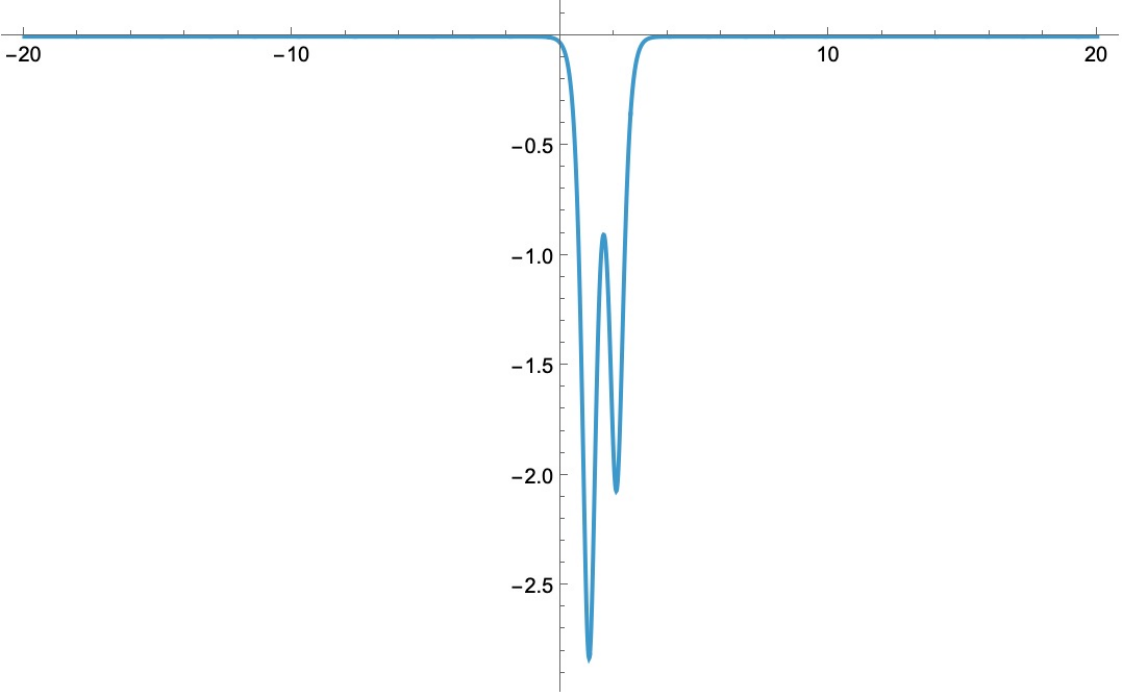}\\
        \includegraphics[width=70mm]{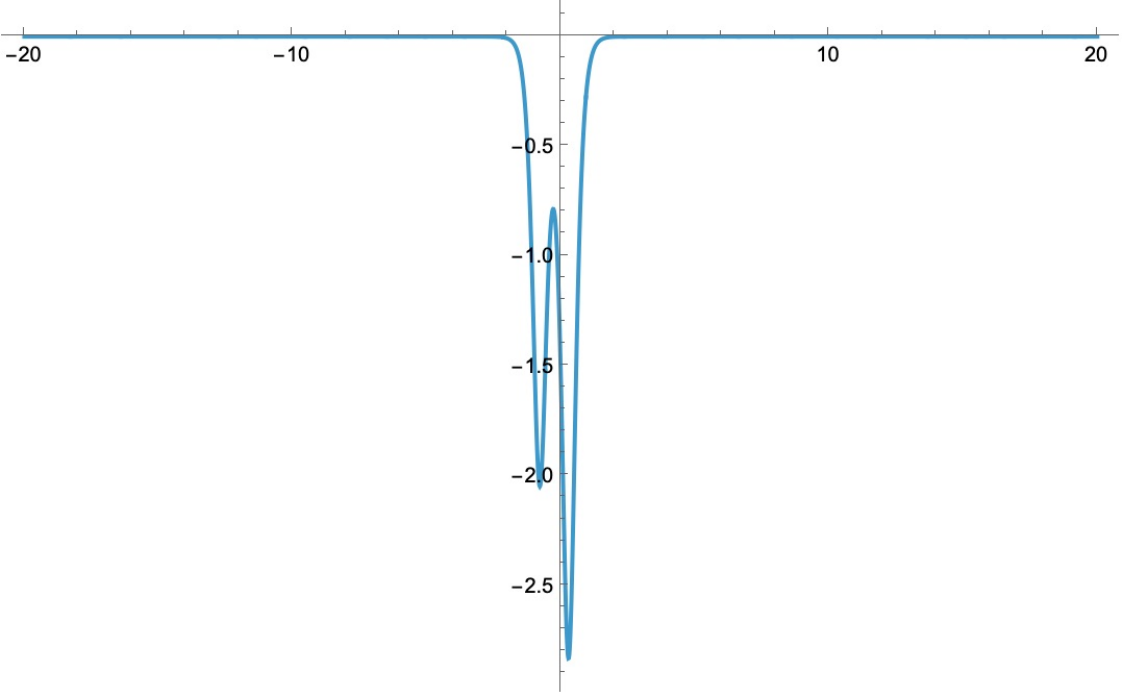}
        \includegraphics[width=70mm]{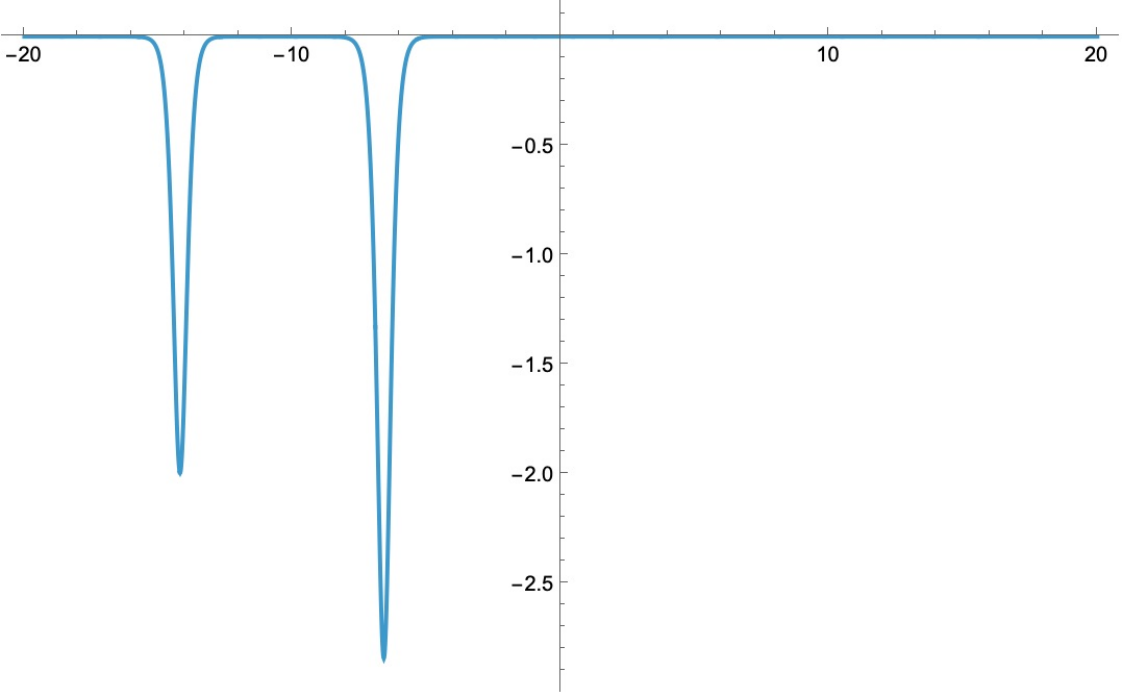}
               \caption{Two-solitons of type DD with $(k_1,k_2)=(5.6,-5.3)$ at $t=2.0$ (upper left), $t=0.3$ (upper right), $t=-0.05$ (lower left), and $t=-2.0$ (lower right).}
        \label{fig:DD1 config}
    \end{center}
\end{figure}

Depending on the depth parameter $\delta$ and the wavenumbers $k_j$, the two-soliton solutions describe EE, DE, and DD collisions. 
Representative examples are shown in Figures~\ref{fig:EE1 config}, \ref{fig:DE1 config}, and \ref{fig:DD1 config}, respectively. 
In all these figures, we set $\delta=1,\;\eta_1=\eta_2=0$.

\bigskip

The same argument extends to the general $N$-soliton solution.
It is known from earlier studies \cite{MatsunoPLA74,ChenPRL43} that the interaction coefficients appearing in the $N$-soliton tau-functions factorise into products of two-body coefficients of the form \eqref{2-soliton interaction}. 
The general tau-functions can be written compactly as
\begin{align}
f^\pm=
\sum_{\mu_1,\ldots,\mu_N\in{0,1}}
\left(
\prod_{j=1}^{N}e^{\mu_jX_j^\pm}
\right)
\left(
\prod_{1\leq i<j\leq N}
A_{ij}^{\mu_i\mu_j}
\right).
\end{align}
In particular, this factorisation satisfies the three-soliton condition for the bilinear equation \cite{Hietarinta1988}.

For example, the three-soliton tau-functions are
\begin{align}\label{tau of 3-soliton}
f^\pm=&
1+e^{X_1^\pm}+e^{X_2^\pm}+e^{X_3^\pm}+
A_{12}e^{X_1^\pm+X_2^\pm}+
A_{23}e^{X_2^\pm+X_3^\pm}+
A_{31}e^{X_3^\pm+X_1^\pm}\nonumber\\
&+A_{123}e^{X_1^\pm+X_2^\pm+X_3^\pm},
\end{align}
where the three-body interaction coefficient factorises as
\begin{align}\label{3-soliton factorisation}
A_{123}=A_{12}A_{23}A_{31}.
\end{align}
Each two-body coefficient \(A_{ij}\) is obtained from
\eqref{2-soliton interaction} by replacing the indices \(1,2\) with
\(i,j\), respectively.

Provided that no zero crosses a contour boundary, the residue index is
\begin{align}\label{N-soliton residue index}
\nu_N=\sum_{j=1}^{N}\nu(k_j),
\end{align}
where $\nu(k)$ is defined by \eqref{one-soliton residue index}, as in the two-soliton cases.

As in the one- and two-soliton cases, an odd-multiple value
\begin{align}
k_j=\frac{(2n+1)\pi}{\delta}
\end{align}
causes zeros of the tau-functions to lie on the real-axis parts of the corresponding contours.
Indeed, as \(k_j\) approaches such a value, the divergence of \(c_j\) implies 
$A_{ij}\longrightarrow1\; (i\neq j)$,
provided that the remaining wavenumbers are admissible.
The $N$-soliton tau-functions therefore factorise as
\begin{align}
f_N^\pm
\longrightarrow
\bigl(1+e^{X_j^\pm}\bigr)
f_{N-1}^\pm.
\end{align}
At the centre of the $j$-th soliton, $e^{X_j^\pm}=e^{\pm i(2n+1)\pi}=-1$, and hence $f_N^\pm=0$.
The logarithmic derivatives
$f_y^\pm/f^\pm$ consequently develop poles on the integration contours, rendering the contour integrals $I^\pm_{\mathrm{in,out}}$ ill defined.

Non-zero even multiples of $\pi/\delta$ must likewise be excluded because the corresponding phase velocity diverges.
In that case, the same factorisation occurs, but the factor $1+e^{X_j^\pm}$ is common to $f^+$ and $f^-$ and cancels from their ratio, so that the physical field degenerates to an $(N-1)$-soliton solution.
For all other admissible wavenumbers, zeros lying strictly inside the contours are retained through the self-consistent residue source $F(x,t)$.

\subsection{Fourier-space analysis of soliton solutions}
\label{subsection:Fourier-space}

In the preceding subsection, we showed that the residue-extended ILW system admits both elevation and depression solitons, with transitions between the two types occurring across the singular values of the soliton wavenumber. 
It is therefore natural to ask how this distinction is manifested in Fourier-space.

In this subsection, we analyse the Fourier spectra of these soliton solutions.
We first derive the Fourier-wavenumber distribution of a one-soliton solution and show that both elevation- and depression-solitons exhibit exponential decay as the magnitude of the Fourier wavenumber increases.
The decay rate is determined by the soliton parameters $k$ and $\delta$, or equivalently by the distance of the nearest complex singularity of the soliton profile from the real axis. 
We then investigate the two-soliton solutions and compare the spectral characteristics of elevation--elevation (EE), depression--elevation (DE), and depression--depression (DD) collisions.
The Fourier-space formulation also provides an independent verification that the non-zero-residue branches satisfy the residue-extended ILW equation, including its self-consistent source term.

Hereafter, we denote the soliton wavenumber by $k$, as in the preceding subsection, and the Fourier wavenumber by $\xi$.

\subsubsection{Fourier spectrum of one-soliton solutions}

We begin with the one-soliton solutions in order to clarify how their spatial profiles are represented in Fourier-space.
As shown in the preceding subsection, the residue-extended ILW system admits both elevation- and depression-solitons, depending on the value of the soliton wavenumber $k$.
Although the two types are described by the same functional form, they belong to different wavenumber ranges, have amplitudes of opposite signs, and lie in different residue sectors. 
This naturally raises the question of how the distinction between elevation- and depression-solitons is manifested in their Fourier spectra.

We focus particularly on the behaviour of the spectrum as the magnitude of the Fourier wavenumber becomes large.
This asymptotic behaviour characterises the spectral localisation of the soliton and, as shown below, is directly related to the analytic structure of its profile in the complex plane. 
In particular, the exponential decay rate is determined by the distance between the real axis and the nearest complex singularity of the soliton profile. 
We shall see that both elevation- and depression-solitons exhibit exponential decay in Fourier-space, while the decay rate depends explicitly on $k$ and $\delta$ and approaches zero as an odd-multiple singular value separating an elevation branch from an adjacent depression branch is approached.

We define the Fourier transform by
\begin{align}\label{Fourier transform}
\widehat{u}(\xi,t)=
\int_{-\infty}^{\infty}
u(x,t)e^{-i\xi x}\,dx.
\end{align}
Since the one-soliton solution is invariant under $k\mapsto-k$, only $|k|$ is relevant.
First consider the initial elevation branch $0<\delta|k|<\pi$, before extending the calculation to the initial depression branch $\pi<\delta|k|<2\pi$.
Substituting the one-soliton profile \eqref{1-soliton} into \eqref{Fourier transform} and using a standard Fourier-integral formula, for example that given in \cite{GradshteynRyzhik1994}, we obtain
\begin{align}\label{FT of elevation 1-soliton}
\widehat{u}(\xi,t)=\widehat{u}_E(\xi,t)=
2\pi
\frac{\sinh(\delta\xi)}
{\sinh(\pi\xi/|k|)}
e^{-i(c(k)t-\eta)\xi}
\end{align}
for the elevation-soliton, $0<\delta\kappa<\pi$, and
\begin{align}\label{FT of depression 1-soliton}
\widehat{u}(\xi,t)=\widehat{u}_D(\xi,t)=
-2\pi
\frac{
\sinh\left(2\pi/|k|-\delta\right)\xi}{\sinh(\pi\xi/|k|)}
e^{-i(c(k)t-\eta)\xi}
\end{align}
for the depression-soliton, $\pi<\delta|k|<2\pi$.
The values at $\xi=0$ in these formulas are understood by continuity.
The higher-branch spectra have the same universal form when expressed in terms of the reduced phase and the scaled Fourier wavenumber $\xi/|k|$; the physical wavenumber $k$ controls only the overall spatial and spectral scales.
The time dependence is entirely contained in a phase factor. 
Consequently, $
\bigl|\widehat{u}(\xi,t)\bigr|$
is independent of time, as expected for a uniformly translating solitary wave.

For the initial elevation branch, the Fourier spectrum has the large-$|\xi|$ asymptotic behaviour
\begin{align}
\bigl|\widehat{u}_{E}(\xi)\bigr|
\sim
e^{-\alpha_{E}}|\xi|,\ 
\alpha_{E}
=
\frac{\pi}{|k|}-\delta>0,
\end{align}
whereas for the initial depression branch,
\begin{align}
\bigl|\widehat{u}_D(\xi)\bigr|
\sim
e^{-\alpha_D|\xi|},\ 
\alpha_D=\delta-\frac{\pi}{|k|}>0.
\end{align}
The two decay rates can therefore be expressed uniformly as
\begin{align}\label{decay rate for 1-soliton}
\alpha
=
\left|
\frac{\pi}{|k|}-\delta
\right|.
\end{align}
For a pair of elevation and depression branches $2m\pi<\delta|k|<(2m+2)\pi,\;m\in\mathbb{Z}_{\geq0}$, the decay rate is given as
\begin{align}\label{one-soliton decay rate}
    \alpha=\left|(2m+1)\frac{\pi}{|k|}-\delta\right|,
\end{align}
This decay rate is precisely the distance from the real axis to the nearest complex singularity of the one-soliton profile.

As $|k|\to\pi/(2m+1)\delta$, the decay rate \eqref{one-soliton decay rate} tends to zero and the Fourier spectrum becomes increasingly broad.
Correspondingly, the soliton profile becomes increasingly localised and approaches a Dirac delta distribution. 
More precisely, with $x_0(t)=c(k)t-\eta$, the limiting profiles satisfy formally
\begin{align}
u_{E}(x,t)
&\longrightarrow
2\pi\delta\bigl(x-x_0(t)\bigr),
\nonumber\\
u_D(x,t)
&\longrightarrow
-2\pi\delta\bigl(x-x_0(t)\bigr),
\end{align}
when the singular value is approached from the elevation and depression sides, respectively.

As an odd-multiple boundary is approached, the nearest zero reaches the real axis, the Fourier decay rate tends to zero, and the residue index changes by one upon crossing the boundary. The broadening of the Fourier spectrum therefore provides a spectral signature of the transition between adjacent residue sectors.

\subsubsection{Fourier spectrum of two-soliton solutions}

We now examine the Fourier spectrum of the two-soliton solution \eqref{2-soliton f} and \eqref{2-soliton interaction}, and show that the interference between the constituent solitons encodes, in a precise and analytically verifiable way, the collision-induced phase shift of the Hirota bilinear formalism, independently of whether each constituent is of elevation- or depression-type.

\begin{figure}
    \centering
    \includegraphics[width=70mm]{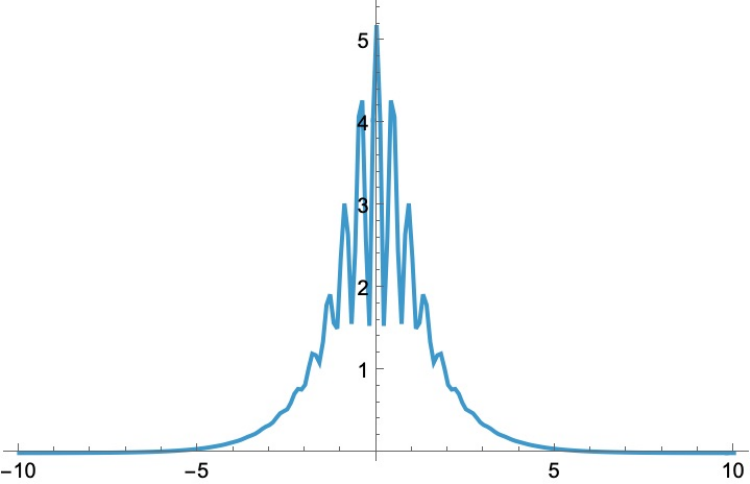}
    \includegraphics[width=70mm]{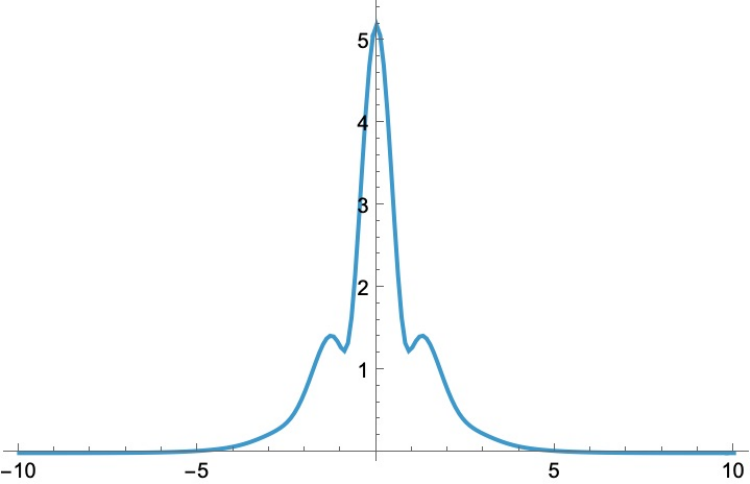}\\
    \includegraphics[width=70mm]{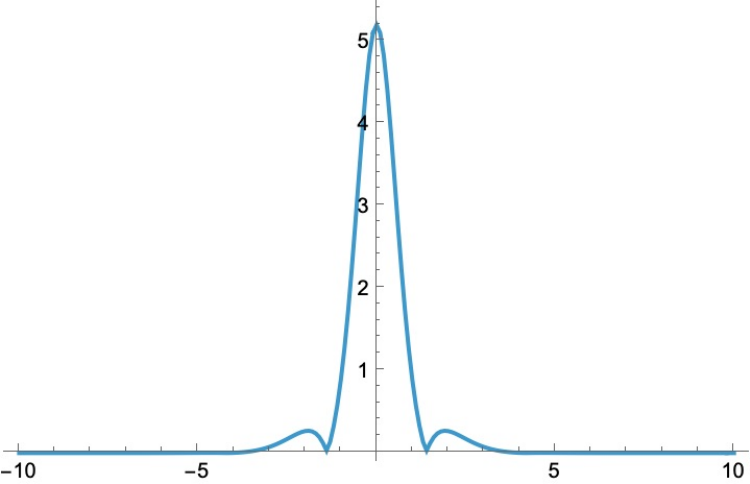}
    \includegraphics[width=70mm]{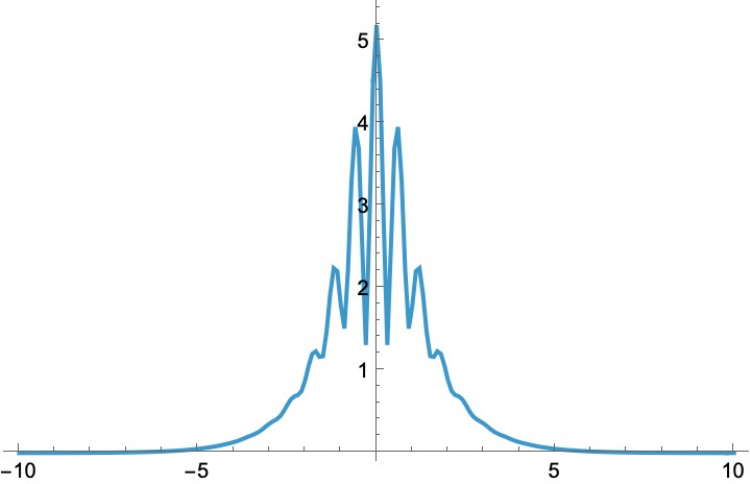}
    \caption{Absolute values of Fourier spectra $|\hat{u}(\xi,t)|$ of two-soliton of type-EE drawn in Figure 4: with $(k_1,k_2)=(1.0,-1.6)$ at $t=15.0$ (upper left), $t=0.0$ (upper right), $t=-2.5$ (lower left), and $t=-15.0$ (lower right).}
    \label{fig:EEFT}
\end{figure}

\begin{figure}
    \centering
    \includegraphics[width=70mm]{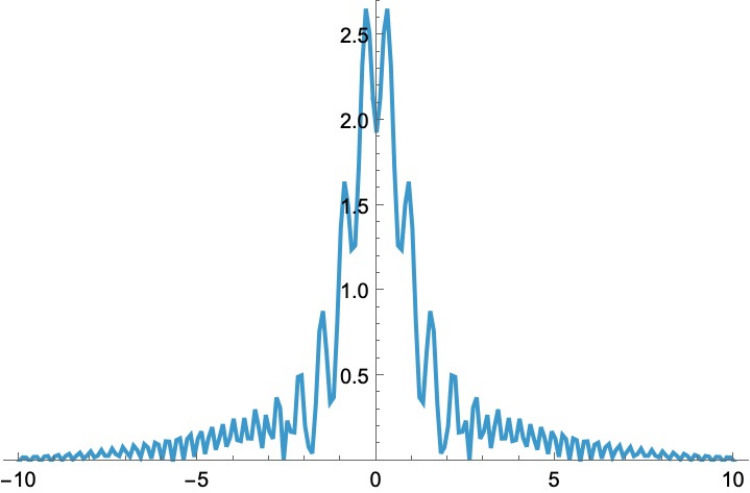}
    \includegraphics[width=70mm]{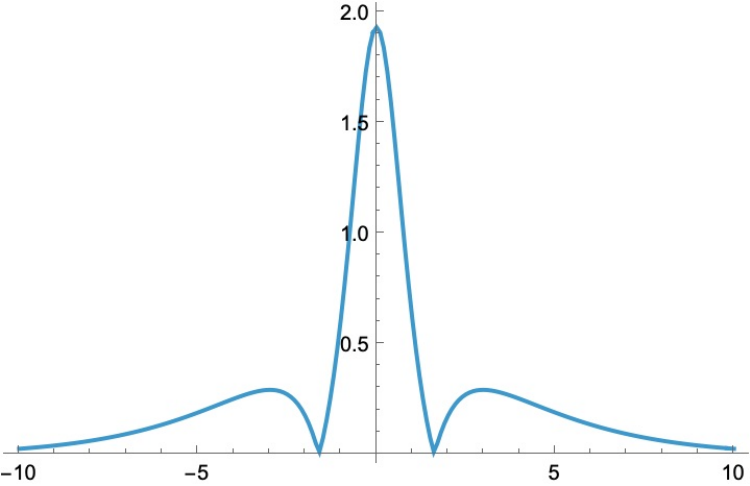}\\
    \includegraphics[width=70mm]{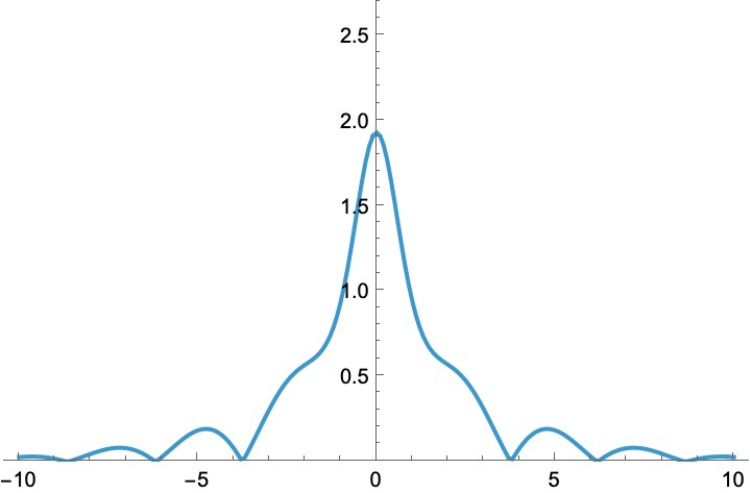}
    \includegraphics[width=70mm]{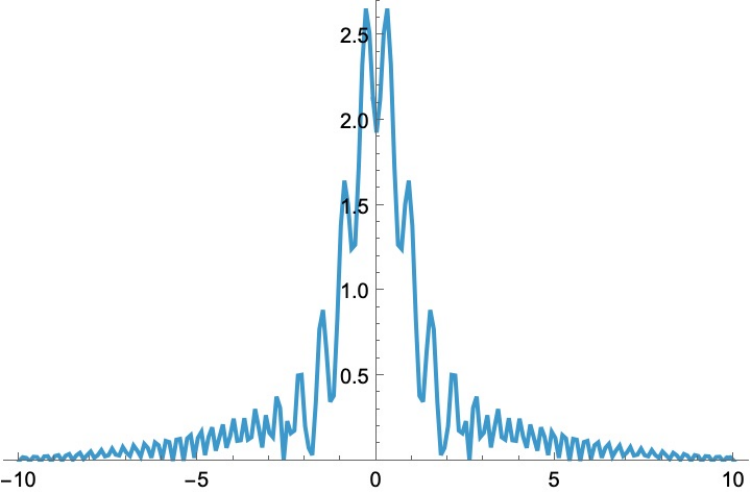}
    \caption{Absolute values of Fourier spectra $|\hat{u}(\xi,t)|$ of two-soliton of type-DE drawn in Figure 5: with $(k_1,k_2)=(1.2,-6.05)$ at $t=0.4$ (upper left), $t=0.0$ (upper right), $t=-0.05$ (lower left), and $t=-0.4$ (lower right).}
    \label{fig:DEFT}
\end{figure}

\begin{figure}
    \centering
    \includegraphics[width=70mm]{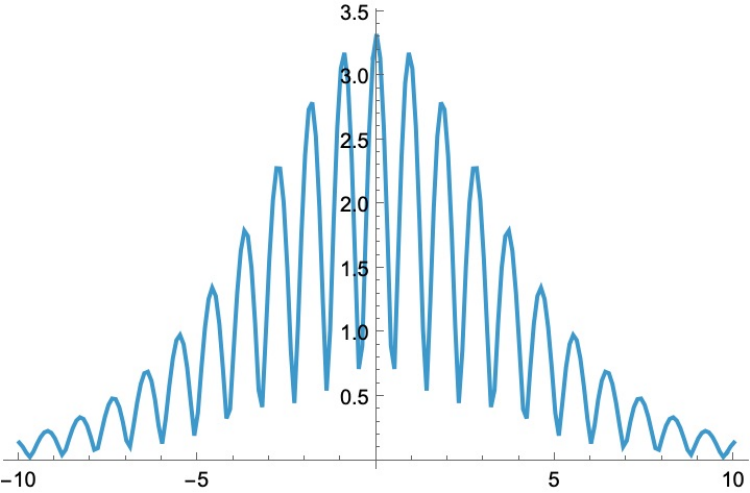}
        \includegraphics[width=70mm]{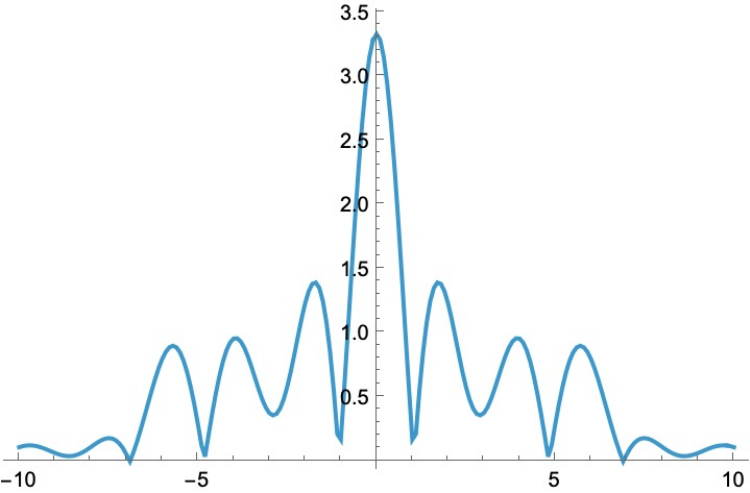}\\
            \includegraphics[width=70mm]{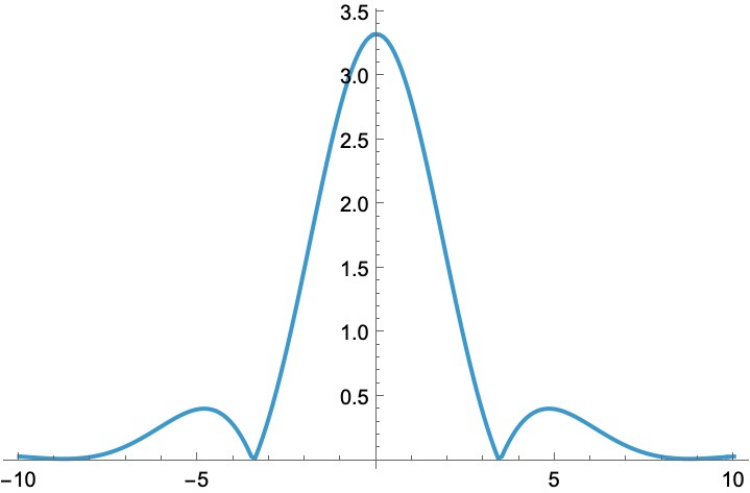}
                \includegraphics[width=70mm]{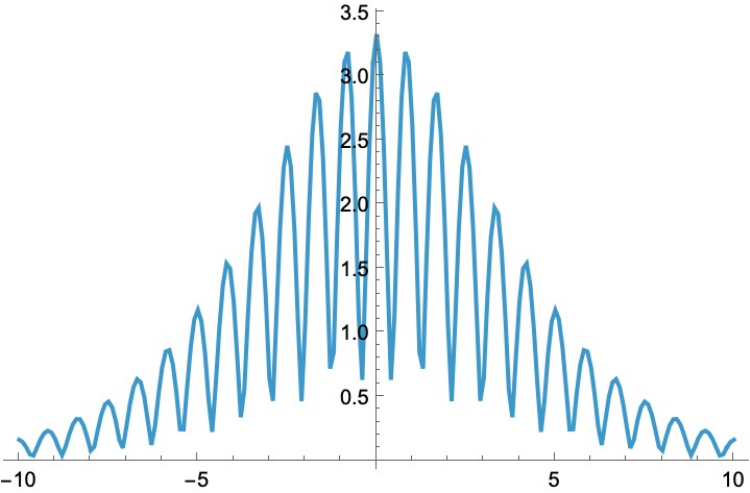}
    \caption{Absolute values of Fourier spectra $|\hat{u}(\xi,t)|$ of two-solitons  of type-DD drawn in Figure 6: with $(k_1,k_2)=(5.6,-5.3)$ at $t=2.0$ (upper left), $t=0.3$ (upper right), $t=-0.05$ (lower left), and $t=-2.0$ (lower right).}
    \label{fig:DDFT}
\end{figure}

Shown below are the Fourier spectra of the two-solitons of several types considered in Section \ref{subsection: two-soliton}.
Those of the type-EE, -DE, and -DD are plotted in Figure \ref{fig:EEFT}, \ref{fig:DEFT}, and \ref{fig:DDFT}, respectively.
We observe that beating periods decrease as $|t|$ increases, i.e., the individual solitons become widely separated.

As is known, this beating is a consequence of the two-body interference, whose oscillation period is approximated to,
\begin{align}\label{beating period}
    \Delta \xi_{\mathrm{beat}}=\frac{2\pi}{|D(t)|},
\end{align}
where $D(t)$ is the phase-shift-corrected separation of each soliton defined below.
We derive \eqref{beating period} and compare it with these numerical analysis of the Fourier spectrum to confirm the consistency of the calculation performed in this study.

First, consider the case of large separation between each soliton, then the two-soliton amplitude is approximated by the superposition of two one-solitons $u_1(x,t)$ and $u_2(x,t)$,
\begin{align}\label{two-soliton large separation}
    u(x,t)\simeq u_1(x,t)+u_2(x,t),
\end{align}
where both elevation and depression cases are possible for $u_1$ and $u_2$ characterised by wavenumbers $k_1$ and $k_2$, respectively.
The peaks of individual solitons $x_1(t)$ and $x_2(t)$ are read from \eqref{1-soliton} as $x_j(t)=c_jt-\eta_j, \;(j=1,2)$. 
The Fourier transform of of \eqref{two-soliton large separation} becomes 
\begin{align}
    \widehat{u}(\xi,t)\simeq \widehat{u}_1(\xi,t)e^{-i(c_1t-\eta_1)\xi}
    +\widehat{u}_2(\xi,t)e^{-i(c_2t-\eta_2)\xi}.
\end{align}
We therefore find that the absolute value of Fourier spectrum is 
\begin{align}
    |\widehat{u}(\xi,t)|^2=|\widehat{u}_1(\xi)|^2+|\widehat{u}_2(\xi)|^2+2|\widehat{u}_1(\xi)||\widehat{u}_2(\xi)|
    \mathrm{Re}\;e^{-i\{(c_1-c_2)t-(\eta_1-\eta_2)\}\xi},
\end{align}
and the beat period $\Delta \xi_{\mathrm{beat}}$ is naively $2\pi/\{(c_1-c_2)t-(\eta_1-\eta_2)\}$, whose denominator is given by the separation width of both solitons $x_1(t)-x_2(t)$.
Note that the Fourier spectra of one-solitons $|\widehat{u}_1|$ and $|\widehat{u}_2|$ determine the envelope and are time-independent, as shown in the previous section.

Next, we show the constants $\eta_1$ and $\eta_2$ must be shifted due to the two-body interference after collision.
Consider the rest frame of the first soliton, the soliton characterised by the wavenumber $k_1$, i.e., $x-c_1t+\eta_1=0$, then we observe $X_2=k_2(x-c_2t+\eta_2)=k_2\{(c_1-c_2)t+\eta_2-\eta_1\}$.
Thus, $X_2\to\infty$ in the region $k_2(c_1-c_2)>0$ as $t\to\infty$ or $k_2(c_1-c_2)<0$ as $t\to-\infty$, we define this region as ``post-collision", so the definition of pre- and post-collision regions are depending on the choice of wavenumber parameters.
At this post-collision region, the tau-function becomes
\begin{align}
    f^\pm=1+e^{X_1^\pm}+e^{X_2^\pm}+A_{12}e^{X_1^\pm +X_2^\pm}\simeq e^{X_2^\pm}(1+A_{12}e^{X_1^\pm}),
\end{align}
and the amplitude tends to
\begin{align}
    u(x,t)=-i\log\frac{f^+}{f^-}&\simeq -i\partial_x\log\frac{1+A_{12}e^{X_1^+}}{1+A_{12}e^{X_1^-}}
    =-i\partial_x\log\frac{1+e^{X_1^++\log A_{12}}}{1+e^{X_1^-+\log A_{12}}}\nonumber\\
    &=-i\partial_x\log\frac{1+\exp\{k_1(x-c_1t+\eta_1+(\log A_{12})/k_1)+i\delta k_1\}}{1+\exp\{k_1(x-c_1t+\eta_1+(\log A_{12})/k_1)-i\delta k_1\}},
\end{align}
by using $e^{X_2^+}/e^{X_2^-}$ is a constant.
This is a one-soliton amplitude with a constant shift $\eta_1\to\eta_1+(\log A_{12})/k_1$, namely, the peak of amplitude shifts by $-(\log A_{12})/k_1$. 
Similarly, considering the rest frame of the second soliton, we find the peak shift of it is $-(\log A_{12})/k_2$.
Hence, we observe that the peak separation of a two-soliton that is sufficiently well-separated is 
\begin{align}\label{peak separation}
    D(t):=(c_1-c_2)t-(\eta_1-\eta_2)-\left(\frac{1}{k_1}-\frac{1}{k_2}\right)\log A_{12},
\end{align}
and the beat period is given as
\begin{align}\label{beat frequency}
    \Delta\xi_{\mathrm{beat}}=\frac{2\pi}{|D(t)|}.
\end{align}
We remark that \eqref{peak separation} applies when $k_1$ and $k_2$ have opposite signs, as in all configurations of the present study.
When these wavenumbers share the same sign, the two solitons' individual shift conditions become mutually exclusive rather than simultaneous, and the shifts are activated one at a time: $D(t)=(c_1-c_2)t-k_1^{-1}\log A_{12}$ on one side of $t=0$ and $D(t)=(c_1-c_2)t+k_2^{-1}\log A_{12}$ on the other, the sides being determined by the sign of $k_2(c_1-c_2)$.

Using the Fourier spectrum and configuration-space plots, we confirm that the phase-shift-corrected separation and beat period formulas, \eqref{peak separation} and \eqref{beat frequency}, hold consistently across all three types of two-solitons, type-EE, -DE, and -DD, as summarised in Table 1, 2, and 3.

In our choice of parameters, $k_2(c_1-c_2)\simeq 1.1$, and $156.9$, for type-EE and type-DE, respectively, both positive, so $t>0$ is the post-collision region for these two types. 
For type-DD, by contrast, $k_2(c_1-c_2)\simeq -17.8$, so $t<0$ is the post-collision region instead.

In most cases, the analytical predictions agree closely with the numerically measured values.
However, near the moment of collision, where the solitons are not sufficiently separated, the approximation becomes less accurate.
The fact that considerable agreement is nevertheless observed in some cases even near $t=0$, where the solitons are almost colliding, can be understood as follows: in the region $|t|\ll1$, the separation \eqref{peak separation} is dominated almost entirely by the constant phase-shift term rather than by $(c_1-c_2)t$, so that the approximation continues to work well provided the phase shift itself is large.

We remark that in the particular DE configuration used here, the correction term $-(k_1^{-1}-k_2^{-1})\log A_{12}$ is numerically very small (since $A_{12}\simeq 1.04$), so that the pre- and post-collision predictions for $D(t)$ agree closely with one another; consequently, the measured separation matches both formulas almost equally well, and this example alone cannot distinguish between the two regimes.
The EE configuration, by contrast, exhibits a large correction ($A_{12}\simeq 8.669$) and was verified at both $t=+15.0$ (post-collision) and $t=-15.0$ (pre-collision), providing the clearest confirmation of the regime-dependent formula.

\begin{table}[htbp]
  \centering
  \label{tab:EE}
  \begin{tabular}{l|rrrr}
 &   post-collision &  &  & pre-collision \\    
 &   $t=15.0$ & $t=0.0$ & $t=-2.5$ & $t=-15.0$ \\
 \hline\\
 Peak separation &  14.0 & 3.6 & 1.8 & 10.5\\
 $|(c_1-c_2)t|$ & 10.3 & 0.0 & 1.7 & 10.3\\
 $D(t)$ & 13.8 & 3.5 & 1.8 & 6.8\\
 \\
 Numerical $\Delta\xi$ &  0.4 & 1.3 & 2.0 & 0.6\\
 $\Delta\xi_{\mathrm{beat}}$& 0.5& 1.8 & 3.5 & 0.9\\
      \end{tabular}
  \caption{Type-EE}

\bigskip\bigskip
  \centering
  \label{tab:DE}
  \begin{tabular}{l|rrrr}
 &   post-collision &  &  & pre-collision \\
 &   $t=0.4$ & $t=0.0$ & $t=-0.05$ & $t=-0.4$ \\
 \hline\\
 Peak separation &  10.4 & 1.3 & 1.4 & 10.4\\
 $|(c_1-c_2)t|$ & 10.4 & 0.0 & 1.3 & 10.4\\
 $|D(t)|$ & 10.4 & 0.0 & 1.3 & 10.3\\
 \\
 Numerical $\Delta\xi$ &  0.7 & 2.9 & 2.3 & 0.7\\
 $\Delta\xi_{\mathrm{beat}}$ & 0.6 & 154.0 & 5.0 & 0.6\\
      \end{tabular}
  \caption{Type-DE}

\bigskip\bigskip

  \centering
  \label{tab:DD}
  \begin{tabular}{l|rrrr}
 &   pre-collision &  &  & post-collision \\    
 &   $t=2.0$ & $t=0.3$ & $t=-0.05$ & $t=-2.0$ \\
 \hline\\
 Peak separation &  6.7 & 1.0 & 1.1 & 7.6\\
 $|(c_1-c_2)t|$ & 6.7 & 1.0 & 0.2 & 6.7\\
 $D(t)$ & 5.8 & 0.1 & 1.1 & 7.6\\
 \\
 Numerical $\Delta\xi$ &  0.9 & 1.7 & 5.0 & 0.8\\
 $\Delta\xi_{\mathrm{beat}}$& 1.1 & 58.1 & 5.9 & 0.8\\
      \end{tabular}
  \caption{Type-DD}

\end{table}

\section{Concluding remarks}\label{section:concluding remarks}

In this paper, we have reconsidered the bilinearisation of the ILW equation from the perspective of Sato hyperfunctions. 
When the tau-functions have no zeros inside the relevant contours, the bilinear equation is equivalent to the conventional source-free ILW equation.
When zeros enter the contours, their contributions must instead be retained as self-consistent residue sources. 
This leads naturally to the residue-extended ILW system, with the conventional ILW equation recovered as its zero-residue sector.

Within this framework, we have constructed explicit elevation- and depression-soliton solutions and classified them according to both their wavenumber ranges and their residue indices. 
The initial elevation branch belongs to the zero-residue sector, whereas the depression and higher elevation branches generally belong to non-zero-residue sectors. 
We have also constructed mixed elevation--depression multisoliton solutions and shown that their residue indices are additive and remain invariant during collisions, provided that no zero crosses a contour boundary. 
The Fourier-space analysis further demonstrated that the exponential decay rate of each soliton spectrum is determined by the distance of the nearest complex singularity from the real axis and tends to zero as an odd-multiple boundary between adjacent residue sectors is approached.

We now consider the deep-water limit $\delta\to\infty$.
As recalled in the Introduction, the source-free ILW equation reduces in this limit to the Benjamin--Ono (BO) equation \cite{SantiniJMP25}
\begin{align}\label{BO equation}
u_t+2uu_x
+
\frac{1}{\pi}
\mathrm{P}\frac{\partial^2}{\partial x^2}
\int_{-\infty}^{\infty}
\frac{u(y,t)}{y-x}\,dy
=0.
\end{align}
The derivation of its bilinear form proceeds analogously to that of the ILW equation, except that the $\tanh$-integral terms appearing in
\eqref{upper part of PV} and \eqref{lower part of PV} disappear in the deep-water limit.
The resulting Hirota bilinear equation has the same form as \eqref{bilinear equation}.

To obtain a non-trivial BO soliton from the initial elevation branch $0<\delta|k|<\pi$ of the ILW solution \eqref{1-soliton}, it is necessary to approach the singular boundary
$\delta|k|=\pi$. 
Let $\kappa=|k|$ and take the double-scaling limit 
\begin{align}\label{elevation BO scaling}
\delta\to\infty,
\
\kappa\to0,
\
\delta\kappa\to\pi-a\kappa+O(\kappa^2),
\end{align}
where $a>0$ is a constant.
In this limit, the phase velocity satisfies
\begin{align}
c(k)=-\kappa\cot(\delta\kappa)
\longrightarrow
\frac{1}{a}.
\end{align}
and we obtain
\begin{align}
f^\pm_{\mathrm{BO},E}
=-k(x-\frac{t}{a}+\eta
\mp ia).
\end{align}
The corresponding field is the standard algebraic elevation soliton of the BO equation \cite{Benjamin1967,OnoJPSJ39}:
\begin{align}\label{BO elevation 1-soliton}
u_{\mathrm{BO},E}(x,t)
=
-i\partial_x
\log\frac{f^+_{\mathrm{BO},E}}{f^-_{\mathrm{BO},E}}
=
\frac{2a}
{\left(x-(t/a)+\eta\right)^2+a^2}.
\end{align}
Thus, the conventional BO soliton is obtained as the deep-water limit of the zero-residue elevation branch of the ILW system.

A corresponding limit can also be taken from the first depression branch $\pi<\delta|k|<2\pi$.
In this case, we set
\begin{align}\label{depression BO scaling}
\delta\to\infty,
\
\kappa\to0,
\
\delta\kappa\to\pi+a\kappa+O(\kappa^2).
\end{align}
The phase velocity then satisfies
\begin{align}
c(k)
=
-\kappa\cot(a\kappa)
\longrightarrow
-\frac{1}{a},
\end{align}
and the limiting tau-functions are
\begin{align}
f^\pm_{\mathrm{BO},D}
=-k(x+\frac{t}{a}+\eta\pm ia).
\end{align}
They yield the negative algebraic profile
\begin{align}\label{BO depression 1-soliton}
u_{\mathrm{BO},D}(x,t)
=
-i\partial_x
\log\frac{f^+_{\mathrm{BO},D}}{f^-_{\mathrm{BO},D}}
=
-\frac{2a}
{\left(x+(t/a)+\eta\right)^2+a^2}.
\end{align}
Note that \eqref{BO depression 1-soliton} is not simply \eqref{BO elevation 1-soliton} with the sign reversed.
Unlike \eqref{BO elevation 1-soliton}, this profile does not belong to the source-free BO sector. 
The complex-conjugate pair $\tilde{y}^\pm=-(t/a)-\eta\mp ia$ of zeros of $f^\pm_{\mathrm{BO},D}$ remains in the corresponding ``strips", i.e., the lower and upper half $y$-plane, respectively, and generates a non-vanishing residue source obtained by taking the limit from \eqref{residue source} as
\begin{align}\label{residue source BO Depression}
    -\frac{\pi}{\delta}\frac{\partial^2}{\partial x^2}F(x,t)\xrightarrow[\delta\to\infty]{}
    &-2\frac{\partial^2}{\partial x^2}\left(\frac{1}{\tilde{y}^+-x}+\frac{1}{\tilde{y}^--x}\right)\nonumber\\
    &=\frac{8\Theta(x,t)(\Theta(x,t)^2-3a^2)}{(\Theta(x,t)^2+a^2)^3},
\end{align}
where $\Theta(x,t)=x+(t/a)+\eta$.
Equation \eqref{BO depression 1-soliton} should therefore be interpreted as a formal depression-soliton of the residue-extended BO system obtained as the deep-water limit of the non-zero-residue ILW sector.

These results suggest several directions for further investigation.
In particular, it remains to formulate the residue-extended BO system explicitly, to determine whether it possesses a Lax pair and an infinite hierarchy of conservation laws, and to examine whether its elevation--depression multisoliton solutions can be embedded in a non-chiral ILW or BO system.
The shallow-water, or KdV, limit also deserves further study.
For fixed $k$, only the initial zero-residue elevation branch survives as $\delta\to0$, while the non-zero-residue sectors require a simultaneous high-wavenumber scaling. 
Understanding this double-scaling regime may clarify whether residue-supported structures have a meaningful counterpart in an extended KdV system.

Finally, clarifying the physical interpretation of the self-consistent residue source may provide a deeper understanding of nonlinear physics and the associated mathematical structures.




\section*{Acknowledgements}
The authors thank all the participants of Soliton Break Through Workshop East 2026 (SBREW-East 2026) for fruitful discussions. 
AN is supported in part by JSPS KAKENHI Grant Number JP23K02794.

\bibliographystyle{iopart-num}

\bibliography{ILWbib}

\end{document}